\documentclass[twocolumn,twocolappendix]{aastex631} % preprint

\usepackage{amsmath}
\usepackage{bm}
\usepackage{CJK}
\usepackage{esint}
\usepackage{float}
\usepackage{graphbox}
\usepackage{makecell}

\graphicspath{{./}{fig/}}

\journalinfo{The Planetary Science Journal, in press} % preprint
\shorttitle{Sodium Brightening of Phaethon}
\shortauthors{Zhang et al.}

\received{2022 October 18}
\revised{2023 March 14}
\accepted{2023 March 27}
\renewcommand{\edit}[2]{#2} % preprint

\begin{document}
\begin{CJK*}{UTF8}{gbsn}

\title{Sodium Brightening of (3200) Phaethon Near Perihelion}

\author[0000-0002-6702-191X]{Qicheng Zhang}
\affiliation{Division of Geological and Planetary Sciences, California Institute of Technology, Pasadena, CA 91125, USA}

\author[0000-0002-8692-6925]{Karl Battams}
\affiliation{US Naval Research Laboratory, 4555 Overlook Avenue, SW, Washington, DC 20375, USA}

\author[0000-0002-4838-7676]{Quanzhi Ye (叶泉志)}
\affiliation{Department of Astronomy, University of Maryland, College Park, MD 20742, USA}
\affiliation{Center for Space Physics, Boston University, 725 Commonwealth Ave, Boston, MA 02215, USA}

\author[0000-0003-2781-6897]{Matthew M. Knight}
\affiliation{Department of Physics, United States Naval Academy, Annapolis, MD 21402, USA}

\author[0000-0002-6917-3458]{Carl A. Schmidt}
\affiliation{Center for Space Physics, Boston University, 725 Commonwealth Ave, Boston, MA 02215, USA}

\correspondingauthor{Qicheng Zhang}
\email{qicheng@cometary.org}

\begin{abstract}
Sunskirting asteroid (3200) Phaethon has been repeatedly observed in STEREO HI1 imagery to anomalously brighten and produce an antisunward tail for a few days near each perihelion passage, phenomena previously attributed to the ejection of micron-sized dust grains. Color imaging by the SOHO LASCO coronagraphs during the 2022~May apparition indicate that the observed brightening and tail development instead capture the release of sodium atoms, which resonantly fluoresce at the 589.0/589.6~nm D lines. While HI1's design bandpass nominally excludes the D lines, filter degradation has substantially increased its D line sensitivity, as quantified by the brightness of Mercury's sodium tail in HI1 imagery. Furthermore, the expected fluorescence efficiency and acceleration of sodium atoms under solar radiation readily reproduce both the photometric and morphological behaviors observed by LASCO and HI1 during the 2022 apparition and the 17 earlier apparitions since 1997. This finding connects Phaethon to the broader population of sunskirting and sungrazing comets observed by SOHO, which often also exhibit bright sodium emission with minimal visible dust, but distinguishes it from other sunskirting asteroids without detectable sodium production under comparable solar heating. These differences may reflect variations in the degree of sodium depletion of near-surface material, and thus the extent and/or timing of any past or present resurfacing activity.
\end{abstract}


\section{Introduction}

Near-Earth asteroid (3200) Phaethon is recognized as the likely parent of the Geminids meteoroid stream \citep[e.g.,][]{whipple1983,davies1984,fox1984,gustafson1989}. Unlike typical icy, cometary meteoroid progenitors, Phaethon has never been observed to present any measurable dust or gas production while near or beyond Earth's orbit \citep[e.g.,][]{hsieh2005,wiegert2008,jewitt2019,ye2021}. While formerly active comets that have exhausted their accessible supply of icy volatiles can appear similarly devoid of activity \citep[e.g.,][]{asher1994,jewitt2005}, Phaethon exhibits higher albedo \citep{green1985}, bluer optical and near-infrared colors \citep{green1985,binzel2001}, and a higher bulk density \citep{hanus2018} than typical of cometary nuclei. Dynamical simulations also indicate that Phaethon most likely originated from the Main Belt like most near-Earth asteroids, rather than the Kuiper Belt or Oort Cloud from which comets are typically sourced \citep[e.g.,][]{bottke2002,deleon2010}. Meanwhile, its dynamical lifetime appears too long for water ice to survive even deep within its interior \citep{jewitt2010}.

Phaethon, however, approaches the Sun more closely than any other known asteroid of its size, to a sunskirting perihelion distance of only $q=0.14$~au \citep{jones2018}. The intense solar heating at such distances can volatilize material much more refractory than water ice to potentially produce comet-like, sublimation-driven activity on even an ice-free Phaethon \citep[e.g.,][]{mann2004,masiero2021,lisse2022}. While its low solar elongation near perihelion thwarts observation of the asteroid by nighttime astronomical facilities during this period, space-borne heliospheric observatories, like the twin Solar Terrestrial Relations Observatory Ahead and Behind (STEREO-A and B) spacecraft \citep{kaiser2008}, do not face this limitation. STEREO-A's Heliospheric Imager 1 (HI1) camera \citep{eyles2009} provided the first reported direct observations of Phaethon near perihelion, which showed that the asteroid underwent a sudden brightening episode during the 2009 apparition that peaked a few hours after perihelion and faded within 2~days \citep{battams2009}.

\citet{jewitt2010} attributed the observed brightening to an impulsive ejection of dust grains from Phaethon's surface, implying that the asteroid is indeed actively losing mass, albeit at a level insufficient to sustain the Geminids stream. The subsequent observable apparitions in 2012 \citep{li2013} and 2016 \citep{hui2017} showed this brightening to be a recurrent phenomenon that repeats with nearly identical timing and amplitude each orbit. \citet{jewitt2013} performed a more careful analysis of the HI1 imagery that further revealed a tail extending a few arcminutes antisunward from Phaethon that develops over the course of $\sim$1~day alongside the brightening, and attributed this tail to micron-sized dust grains being rapidly accelerated and swiftly dispersed by solar radiation pressure. Such grains would constitute a class of dust distinct from the millimeter-sized and larger grains still orbiting alongside Phaethon as part of the Geminids meteoroid stream.

Our current analyses, however, show that the observed brightening and tail cannot be attributed to dust grains of any form. In the following sections, we present additional observations of Phaethon's activity collected by the HI1 cameras of both STEREO-A and B, as well as by the Large Angle Spectrometric Coronagraph (LASCO) C2 and C3 coronagraphs \citep{brueckner1995} onboard the Solar and Heliospheric Observatory (SOHO) spacecraft \citep{domingo1995}. We use the morphology and photometry provided by these observations to demonstrate that Phaethon's observed brightening and tail actually capture resonance fluorescence by atomic sodium (henceforth, \ion{Na}{1}) liberated under solar heating, compare this behavior with that of several other sunskirting objects, and discuss the associated implications on the formation of the Geminids stream.

\section{Data}

\begin{deluxetable}{lll}
\tablecaption{Observable Apparitions of (3200) Phaethon}
\label{tab:app}

\tablecolumns{3}
\tablehead{
\colhead{Apparition $T_\mathrm{p}$} & \colhead{Instrument} & \colhead{Observable Period} \\
\colhead{(UT)\tablenotemark{a}} & \colhead{} & \colhead{(UT)\tablenotemark{b}}
}

\startdata
1996 Jul 25 23:49 & SOHO LASCO C3 & Jul 24--27\tablenotemark{c} \\
1997 Dec 31 13:25 & SOHO LASCO C3 & Dec 30--1998 Jan 02 \\
1999 Jun 08 03:21 & SOHO LASCO C2 & Jun 08--09 \\
& SOHO LASCO C3 & Jun 06--08\tablenotemark{c} \\
2000 Nov 12 20:19 & SOHO LASCO C2 & Nov 12--13 \\
& SOHO LASCO C3 & Nov 11--14 \\
2002 Apr 20 09:52 & SOHO LASCO C3 & Apr 18--21 \\
2003 Sep 25 22:55 & SOHO LASCO C3 & Sep 24--27 \\
2005 Mar 02 13:04 & SOHO LASCO C3 & Feb 28--Mar 02 \\
2006 Aug 08 04:13 & SOHO LASCO C3 & Aug 06--09 \\
2008 Jan 13 18:54 & SOHO LASCO C3 & Jan 13--15 \\
& STEREO-B HI1 & Jan 03--13 \\
2009 Jun 20 07:21 & SOHO LASCO C2 & Jun 21--22 \\
& SOHO LASCO C3 & Jun 19--21 \\
& STEREO-A HI1 & Jun 17--22 \\
& STEREO-B HI1 & Jun 21--30 \\
2010 Nov 25 17:51 & SOHO LASCO C2 & Nov 25--26 \\
& SOHO LASCO C3 & Nov 24--27 \\
& STEREO-B HI1 & Nov 15--19\tablenotemark{c} \\
2012 May 02 07:48 & SOHO LASCO C3 & Apr 30--May 03 \\
& STEREO-A HI1 & Apr 30--May 04 \\
& STEREO-B HI1 & Apr 22--May 03 \\
2013 Oct 07 21:18 & SOHO LASCO C3 & Oct 06--09 \\
& STEREO-B HI1 & Oct 09--18\tablenotemark{c} \\
2015 Mar 15 07:47 & SOHO LASCO C3 & Mar 13--15 \\
2016 Aug 19 19:45 & SOHO LASCO C3 & Aug 18--20 \\
& STEREO-A HI1 & Aug 18--30 \\
2018 Jan 25 08:24 & SOHO LASCO C3 & Jan 25--27 \\
2019 Jul 02 23:59 & SOHO LASCO C3 & Jul 02--04 \\
& STEREO-A HI1 & Jul 03--13 \\
2020 Dec 07 14:35 & SOHO LASCO C2 & Dec 08--09 \\
& SOHO LASCO C3 & Dec 06--09 \\
2022 May 15 04:21 & SOHO LASCO C2 & May 15 \\
& SOHO LASCO C3 & May 13--17\tablenotemark{d} \\
& STEREO-A HI1 & May 15--25\tablenotemark{e}
\enddata

\tablenotetext{a}{Time of perihelion for apparition.}
\tablenotetext{b}{Period with Phaethon observable in the field of C2/C3 at $r<0.16$~au, or in the field of HI1 at $r<0.4$~au (but only periods containing $r<0.16$~au are listed).}
\tablenotetext{c}{Insufficient sensitivity for clear detection of activity.}
\tablenotetext{d}{Only detected in special May 15--16 Orange observations.}
\tablenotetext{e}{Only May 15--20 included in analysis.}
\end{deluxetable}

Phaethon's perihelic activity has previously only ever been reported to be seen by STEREO-A HI1, with detailed analyses published of the 2009, 2012, and 2016 apparitions as discussed above. Phaethon, however, should also have been observable by STEREO-B HI1 near perihelion in several apparitions, and crosses the field of SOHO LASCO C3, if not also C2, near perihelion in every apparition. We anticipated the 2022 apparition to be particularly favorable for LASCO, with Phaethon crossing the C2 field $\sim$0.5~day after perihelion at the expected peak of its brightening, and carried out a special observing sequence with both coronagraphs to better characterize the activity with higher spatial resolution through photometric bandpasses different from that of previously analyzed HI1 data. Below, we discuss our analysis of both the new and archival data, which collectively capture Phaethon's activity in all 18 apparitions from 1997 to 2022, as summarized in Table~\ref{tab:app}.

\begin{figure*}
\includegraphics[width=\linewidth]{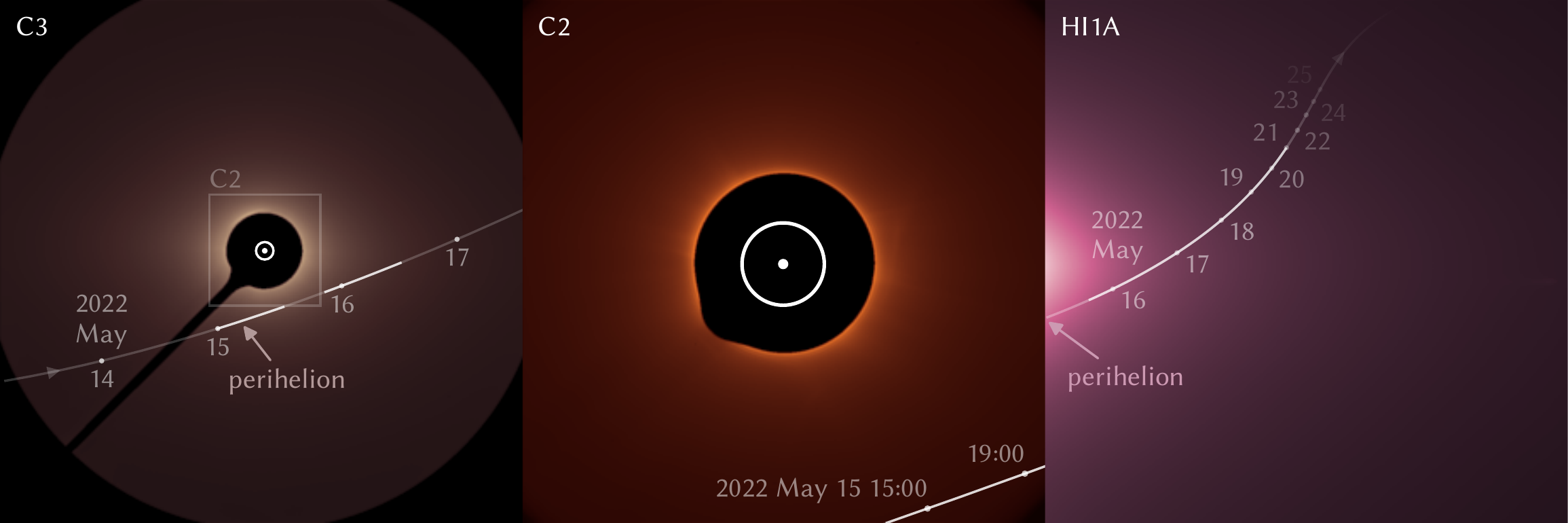}
\caption{Apparent trajectory of (3200) Phaethon during its 2022 apparition through the fields of view of the SOHO LASCO C3 (left) and C2 (middle) coronagraphs, and of the STEREO-A HI1 imager (HI1A; right), with dates and the location of perihelion labeled. Highlighted sections of the track indicate periods of each camera's observations included in our analysis of this apparition. The $\odot$ markers on the C3 and C2 panels indicate the size and position of the Sun behind each camera's occulter, while the square on the C3 panel indicates the relative field of view of C2. The Sun is off the left edge of the HI1A frame.}
\label{fig:traject}
\end{figure*}

\begin{deluxetable*}{lllccc}
\tablecaption{Observations from the 2022 Apparition of (3200) Phaethon}
\label{tab:2022obs}

\tablecolumns{6}
\tablehead{
\colhead{Instrument} & \colhead{Exposures} & \colhead{Observation Time} & \colhead{$r$} & \colhead{$\varDelta$} & \colhead{$\alpha$} \\
\colhead{} & \colhead{} & \colhead{(UT)} & \colhead{(au)\tablenotemark{a}} & \colhead{(au)\tablenotemark{b}} & \colhead{($^\circ$)\tablenotemark{c}}
}

\renewcommand\cellalign{tr}
\startdata
SOHO LASCO C3 & $166\times60$~s Orange & \makecell{2022 May 15 00:00--12:41, \\ 23:03--May 16 11:57\phantom{,}} & 0.140--0.151 & 0.863--0.868 & 151.2--167.1 \\
& $19\times120$~s Blue & 2022 May 15 20:39--22:57 & 0.143--0.144 & 0.863 & 162.7--164.3 \\
SOHO LASCO C2 & $30\times60$~s Orange & \makecell{2022 May 15 13:14--15:07, \\ 17:42--20:03\phantom{,}} & 0.141--0.143 & 0.863 & 164.4--167.1 \\
& $15\times120$~s Blue & 2022 May 15 15:16--17:35 & 0.142 & 0.863 & 166.0--166.8 \\
STEREO-A HI1\tablenotemark{d} & $152\times1200$~s & \makecell{2022 May 15 16:23\phantom{--00:00,} \\ --May 20 23:53\phantom{--00:00,}} & 0.142--0.270 & 0.856--1.067 & 60.4--135.7
\enddata

\tablenotetext{a}{Heliocentric distance.}
\tablenotetext{b}{Distance from observer.}
\tablenotetext{c}{Phase angle.}
\tablenotetext{d}{Data from standard, synoptic observing program.}
\end{deluxetable*}
\vspace{-2.5em}

\begin{figure*}
\centering
\includegraphics[width=0.8\linewidth]{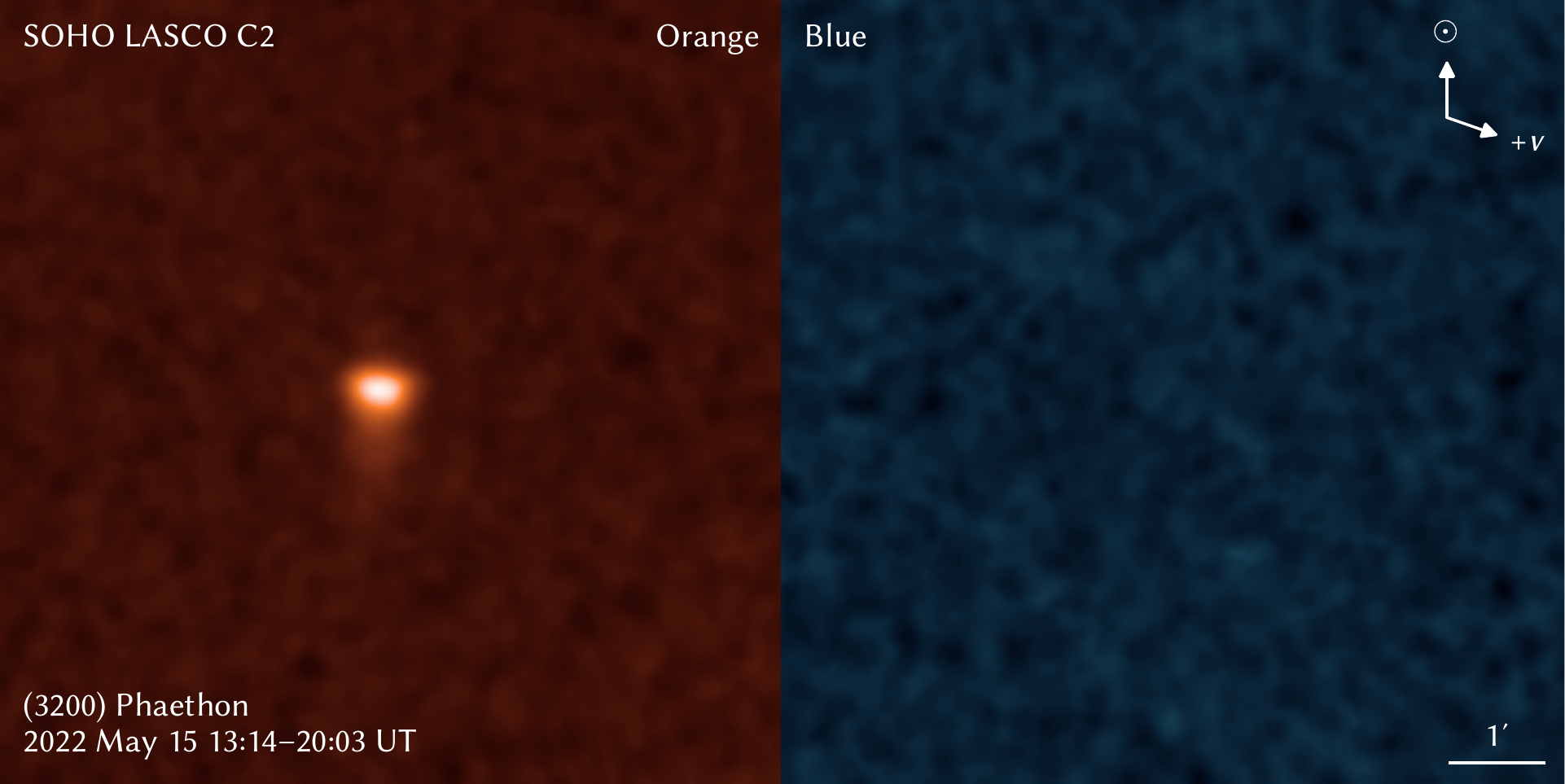}
\hspace*{0.005\linewidth}\includegraphics[width=0.53\linewidth]{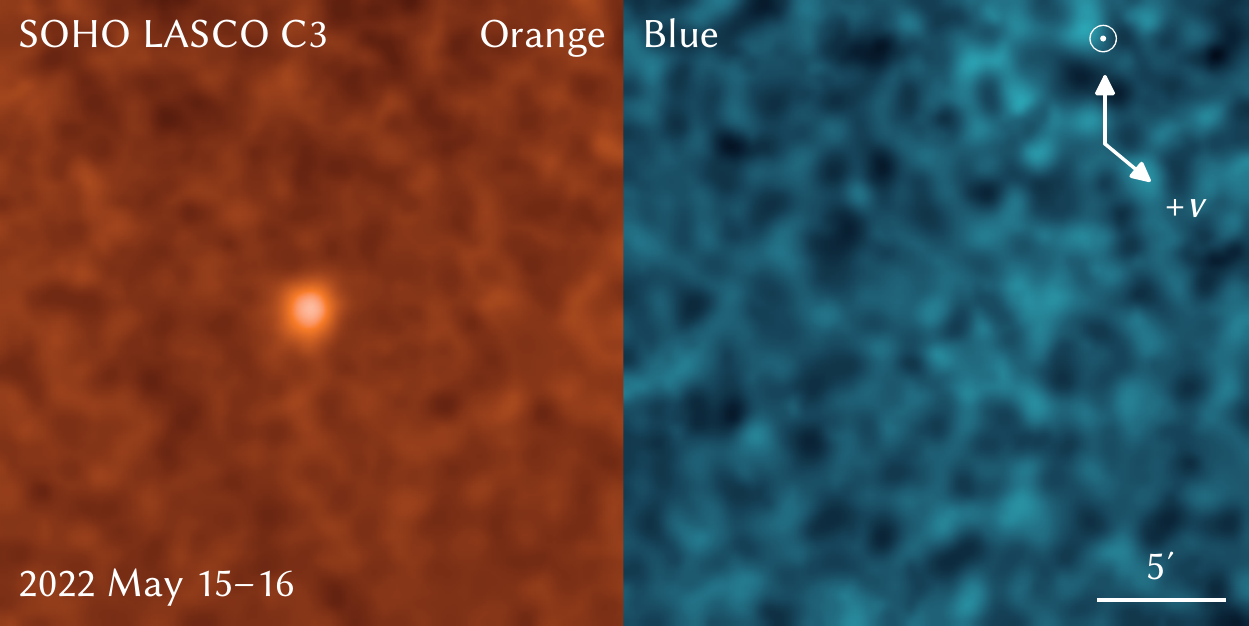}\hspace{0.005\linewidth}\includegraphics[width=0.265\linewidth]{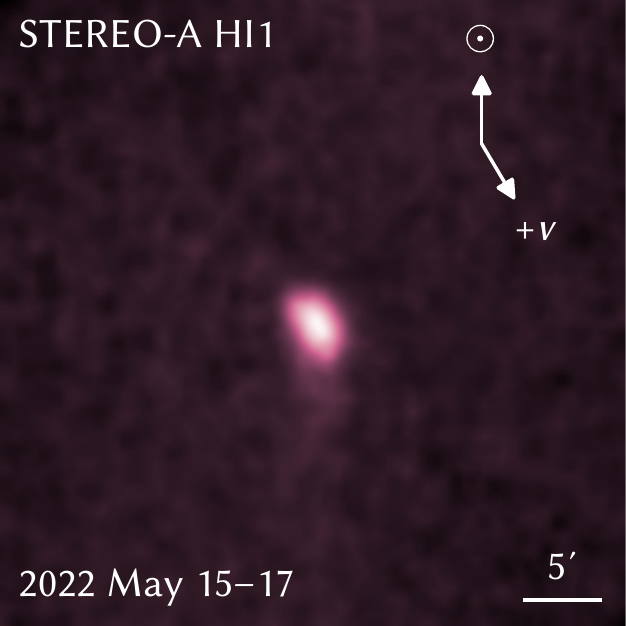}
\caption{Stacked frames of Phaethon during its 2022 apparition from SOHO LASCO C2 (top), LASCO C3 (lower left), and STEREO-A HI1 (lower right). Phaethon and its tail are much brighter through the Orange filters, which transmit \ion{Na}{1} D emission, than through the Blue filters, which block \ion{Na}{1} D emission and where Phaethon is not visible. Each Orange/Blue pair has been scaled such that a solar-colored source would appear similarly bright in both frames to permit visual comparison. The upward $\odot$ arrows indicate the sunward direction, to which individual frames were aligned prior to stacking, while the $+v$ arrows indicate the heliocentric velocity direction near the midpoint time of each stack. The C2 and HI1 stacks show slight trailing due to the motion of Phaethon over individual exposures together with small errors in the astrometric solutions used to align the individual frames.}
\label{fig:stack2022}
\end{figure*}

\subsection{Instruments}

\subsubsection{STEREO-A/B HI1}

The STEREO mission involves two, functionally identical spacecraft on heliocentric orbits similar to that of Earth \citep{kaiser2008}: STEREO-A, which orbits slightly interior to and faster than Earth, and STEREO-B, which orbits slightly exterior to and slower than Earth. Their slight orbital differences cause them to separate, drifting from Earth at ${\sim}22$~deg~yr$^{-1}$. Since the commencement of standard operations in 2007~January, STEREO-A and B have drifted nearly one complete orbit ahead and behind to return to the vicinity of Earth in 2023, although STEREO-B ceased operations in 2014 after its loss of contact \citep{ossing2018}.

The Sun Earth Connection Coronal and Heliospheric Investigation (SECCHI) instrument suite \citep{howard2008} onboard each spacecraft contains a pair of coronagraphs (COR1/2) and heliospheric imagers (HI1/2). Our analysis covers only the HI1 instruments, the inner heliospheric imagers targeting the region of sky between $4^\circ$ and $24^\circ$ from the Sun along the Sun--Earth line---fields well-suited to observing Phaethon near perihelion. Each camera has an unbinned image scale of 36~arcsec~px$^{-1}$ \citep{brown2009}. Under its standard, synoptic observing program, HI1 typically returns 36 onboard-processed frames per day, each of which is a sum of $30\times40$~s exposures binned $2\times2$ to 72~arcsec~px$^{-1}$. For our analysis, we used the standard level 2 data, which provide astrometric calibrations and eliminate the gradient of the F-corona through subtraction of a background frame calculated as the mean of the lowest quartile of each pixel across all frames over a 1~day window. However, we used the updated photometric calibrations by \citet{tappin2017} and \citet{tappin2022} in place of the standard level 2 photometric calibrations. To improve sensitivity, we also subtracted field stars from each frame, using a sidereally-aligned median stack of neighboring frames for the stellar background model.

Each HI1 camera observes through a fixed bandpass filter, which preflight calibrations showed had full width at half maximum (FWHM) wavelength spans of $\sim$615--740~nm, with a blue leak near 400~nm and a red leak near 1000~nm \citep{bewsher2010}. The filters effectively blocked the 589.0/589.6~nm \ion{Na}{1} D resonance lines with a relative transmission of only 1--2\%, which led \citet{jewitt2013} and \citet{hui2017} to rule out \ion{Na}{1} D emission as the source of Phaethon's brightening in HI1. However, the appearance of Mercury's \ion{Na}{1} tail several times brighter than expected in HI1 called the preflight filter measurements into question \citep{schmidt2010b}. \citet{halain2012} subsequently re-evaluated the filter transmission on the HI1 engineering qualification model, which revealed the true HI1 bandpass to actually be blueshifted from preflight values by $\sim$20~nm, raising the relative \ion{Na}{1} D transmission to $\sim$15\%, roughly in line with observations. The MESSENGER mission has shown that Mercury's Na escape rate varies in a seasonally repeating pattern \citep{cassidy2021}. This property enables us to use the tail as a flux standard to calibrate HI1's sensitivity to \ion{Na}{1} D emission, as we present in Appendix~\ref{sec:hi1sens}.

\subsubsection{SOHO LASCO C2/C3}

Unlike STEREO, SOHO monitors the Sun and the heliospheric environment from the vicinity of Earth in a halo orbit around the Sun--Earth L1 point. Its LASCO C2 and C3 coronagraphs have observed thousands of other objects active near the Sun, and unlike HI1, each contain a set of interchangeable bandpass filters that can measure the colors of observation targets \citep{battams2017}. LASCO C2 covers a narrow region 1.5--6~$R_\odot$ above the solar limb at 12~arcsec~px$^{-1}$, while LASCO C3 observes a wider region 3.7--30~$R_\odot$ from the limb at a coarser 56~arcsec~px$^{-1}$. Under their standard, synoptic program, C2 and C3 generally alternate in observations, presently with a 12~minute interval between images from the same camera. C2 records 25~s exposures through a $\sim$540--620~nm (FWHM span) Orange filter while C3 records 18~s exposures through a $\sim$530--840~nm Clear filter, with only sporadic exposures through other filters \citep{battams2017}. Special observing sequences can be scheduled several days in advance to use other filter and exposure combinations.

For our analyses, we began with the minimally processed level 0.5 data and applied the equivalent of level 1 bias and vignetting corrections \citep{thernisien2003}. We also removed the coronal gradient by subtracting the median of exposure-normalized frames with the same camera/filter combination in each apparition as a background frame. To avoid further degradation to LASCO's already undersampled point spread functions (PSFs), we skipped the level 1 stage that corrects image distortions by interpolation onto an undistorted grid. We instead incorporated the supplied radial polynomial distortion coefficients into our astrometric solutions, which we then fitted to stars in the Gaia DR3 catalog \citep{gaia2022}. Visual inspection suggests these solutions are accurate to $<$1~px over roughly half the C2 field, and over all but the inner and outermost few percent of the C3 field. We also performed new photometric calibrations of C2 and C3 that span SOHO's lifetime in order to more confidently constrain potential variations in the sensitivity of the cameras over time. We present these calibrations in Appendix~\ref{sec:lascophot}.

\subsection{Observations}

\begin{figure*}
\centering
\includegraphics[width=\linewidth]{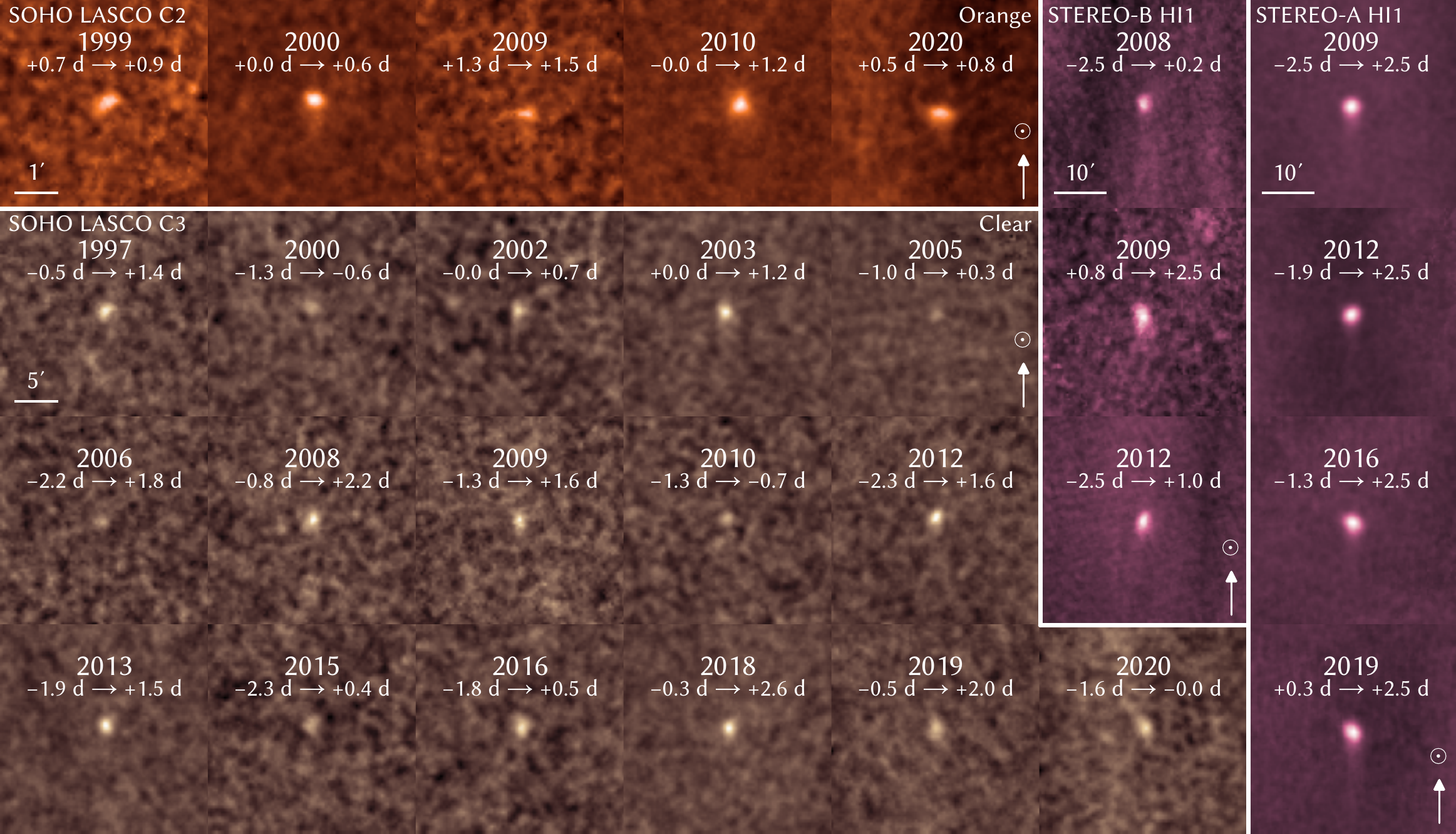}
\caption{Phaethon's perihelion activity \edit{1}{at all 17 apparitions} between 1997 and 2020, as observed by SOHO LASCO C2/C3 and by both STEREO-A and STEREO-B HI1. \edit{1}{Text labels in each image indicate the perihelion year of the respective apparition along with the time spanned by the image stack, given in days relative to the perihelion time.} The sunward direction is oriented upward in all frames.}
\label{fig:appar}
\end{figure*}

\subsubsection{2022 Apparition}
\label{sec:2022app}

During its 2022 apparition, Phaethon reached perihelion at 2022~May~15 04:21 UT (JPL orbit solution 777). It crossed the LASCO C3 field of view over May 14--17 and the C2 field of view on May~15 13--20h UT, with the latter interval coinciding with the timing of Phaethon's previously reported HI1 brightening peak $\sim$0.5~days after perihelion. We conducted a special sequence of color observations to characterize Phaethon with LASCO within this period. Phaethon also entered the STEREO-A HI1 field of view on May~15 where it was concurrently monitored by the standard, synoptic observing program alongside our special LASCO C2 and C3 observations, and for the remainder of the apparition. Figure~\ref{fig:traject} illustrates Phaethon's trajectory through the fields of all three cameras.

Typical sunskirting comets tend to appear much brighter in Orange than in Clear-filtered observations, often by $\gtrsim$1~mag relative to solar color, due to intense \ion{Na}{1} fluorescence that outshines the sunlight scattered by dust \citep[e.g.,][]{biesecker2002,knight2010,lamy2013}. Both the Orange and Clear filters strongly transmit the \ion{Na}{1} D lines, but the much narrower Orange bandpass preferentially transmits \ion{Na}{1} D emission relative to light with Sun-like spectra, leading pure \ion{Na}{1} D emission to appear $\sim$1.3~mag brighter in Orange than in Clear data, as calculated in Appendix~\ref{sec:lascophot}. We considered that Phaethon may behave similarly, so to improve LASCO's sensitivity to Phaethon's potential \ion{Na}{1} activity, we scheduled a sequence of mainly 60~s Orange exposures by both C2 and C3 spanning 2022~May~15 0h and May~16 12h, with only C2 observations during its May~15 13--20h window, and C3 observations filling the remaining time. We also scheduled blocks of 120~s exposures through the $\sim$460--515~nm C2 and C3 Blue filters, which effectively block \ion{Na}{1} D emission at ${<}1\%$ relative transmission, to constrain the presence of any micron-sized dust associated with the activity. Table~\ref{tab:2022obs} summarizes these observations.

The plan immediately proved successful, with Phaethon appearing so bright in our C2 Orange frames that it was unwittingly noticed and reported to the Sungrazer citizen science project as a new comet by Zhijian Xu. Closer inspection also revealed it to be visible in the C3 Orange sequence, but not in either of the Blue sequences. Figure~\ref{fig:stack2022} compares the $2.5\sigma$-clipped, JPL Horizons ephemeris-aligned stacks of all Orange and Blue frames from each camera, illustrating Phaethon's activity to be prominent in Orange at apparent magnitude 8.8 in C2 within a $45''$ radius aperture, yet not at all in Blue to a $3\sigma$ limiting magnitude of 10.9.

Thanks to the strong forward scattering enhancement at its high phase angle of $\alpha\approx166^\circ$, the C2 Blue limit translates to a stringent absolute magnitude limit of $H>20.2$ for any comet-like, micron-sized dust following the Schleicher--Marcus phase function \citep{schleicher2011,marcus2007}, corresponding to a cross section of ${\lesssim}0.1$~km$^2$ at Phaethon's $\sim$11\% geometric albedo \edit{1}{over visible wavelengths \citep[i.e., including the C2 Blue bandpass;][]{binzel2001,maclennan2022}}, or $\lesssim$400~kg of 1~$\mu$m radius grains with a Geminids-like density of 2.6~g~cm$^{-3}$ \citep{borovivcka2010}---three orders of magnitude below \citet{jewitt2013}'s estimate. Motivated by our findings, \citet{hui2023} subsequently performed a similar analysis of Phaethon's lack of forward scattering in archival data from STEREO's COR2 coronagraphs, which are also insensitive to \ion{Na}{1} D emission, and arrived at constraints comparable to our result. These initial findings validate our suspicion that Phaethon's activity is observationally similar to that of typical sunskirting comets seen by LASCO, and constitutes our first line of evidence attributing the observed brightening to \ion{Na}{1} D emission rather than dust.

\subsubsection{Earlier Apparitions}

Encouraged by the prominence of Phaethon in our 2022 Orange data, we revisited all of the C2 Orange and C3 Clear data collected by the synoptic program over SOHO's operating lifetime where Phaethon was within the frame at $r<0.16$~au, and produced similar ephemeris-aligned stacks to evaluate Phaethon's activity across apparitions. We repeated the process with both STEREO-A and B HI1 for completeness. In doing so, we recovered Phaethon in LASCO C2 and/or C3 at all 18 apparitions since 1997 as well as at several additional apparitions in HI1, as shown in Figure~\ref{fig:appar}. Sensitivity varies between apparitions due to differences in the viewing geometry and track Phaethon takes across each camera's field of view at every apparition, which affects the both the time span of observations and the level of noise from coronal background, the latter of which increases at lower elongations. Only the 1996 apparition lacks a clear detection due to the low cadence of LASCO data at the commencement of SOHO operations. As we demonstrate in the next section, Phaethon appears brighter in these near-perihelion detections than expected from sunlight scattered by its solid surface alone, indicating they capture the sizable brightness contribution of its \ion{Na}{1} activity.

\section{Analysis}

\begin{figure*}
\centering
\includegraphics[width=\linewidth]{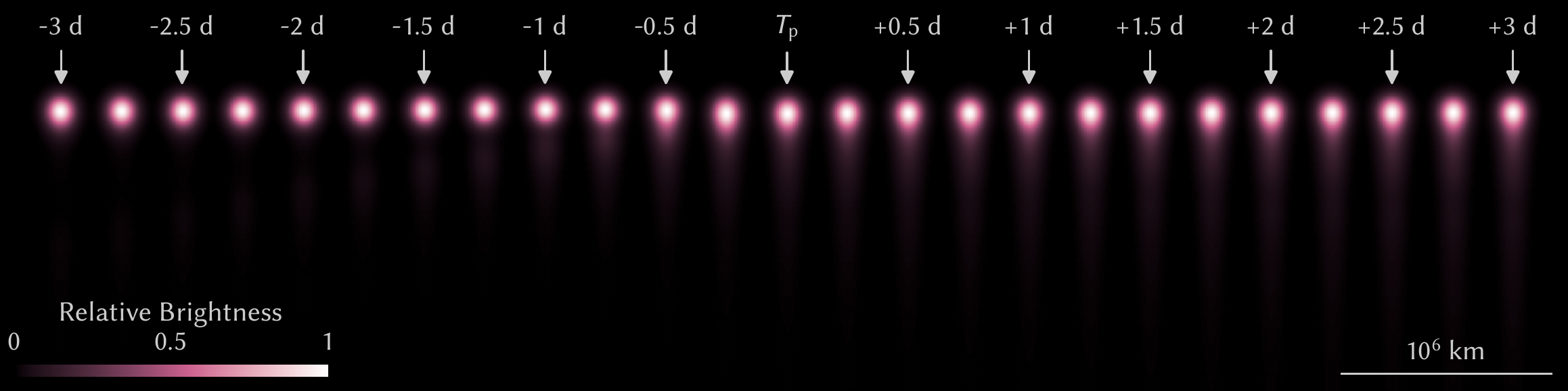}
\includegraphics[width=0.49\linewidth]{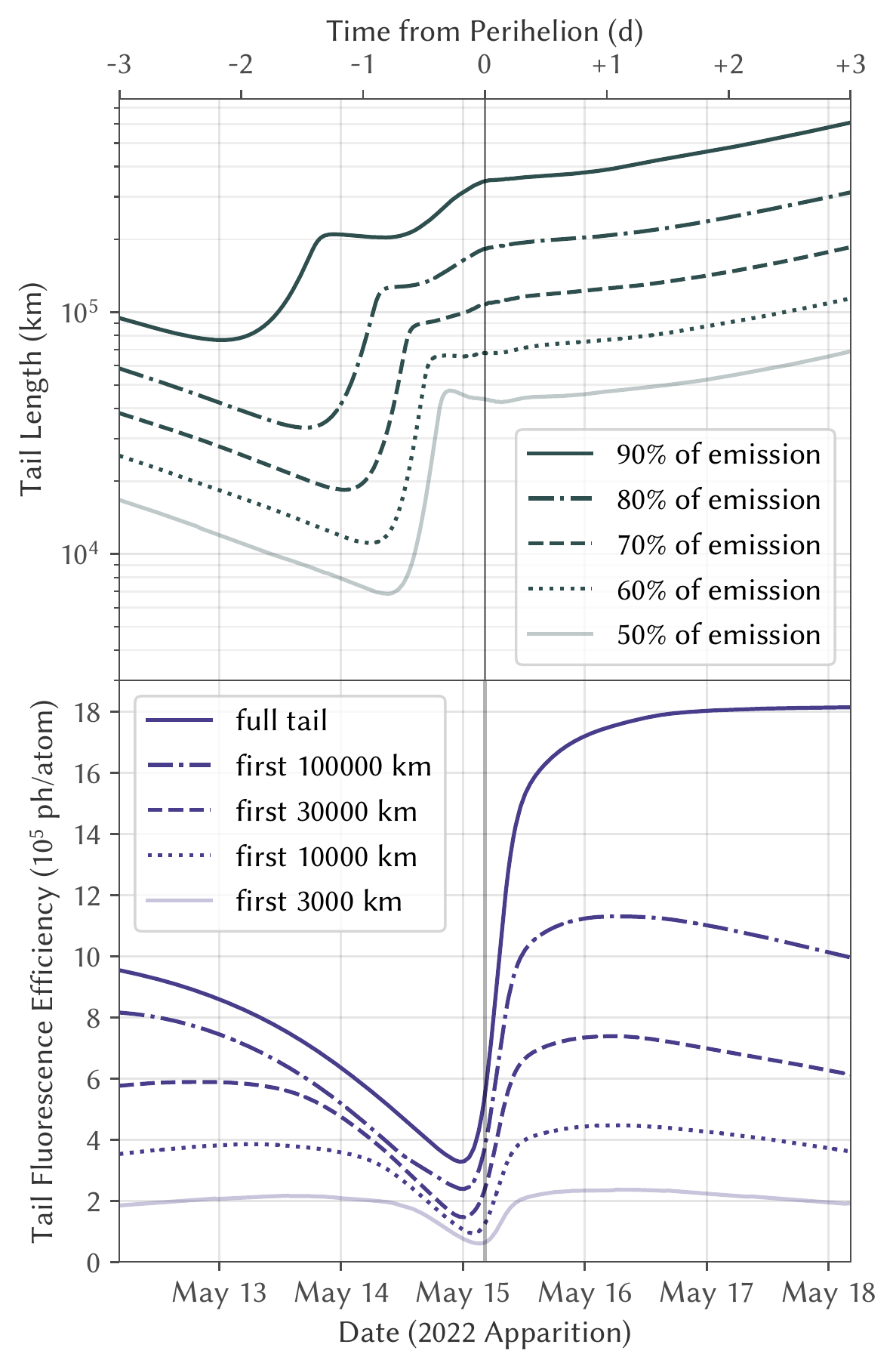}
\includegraphics[width=0.49\linewidth]{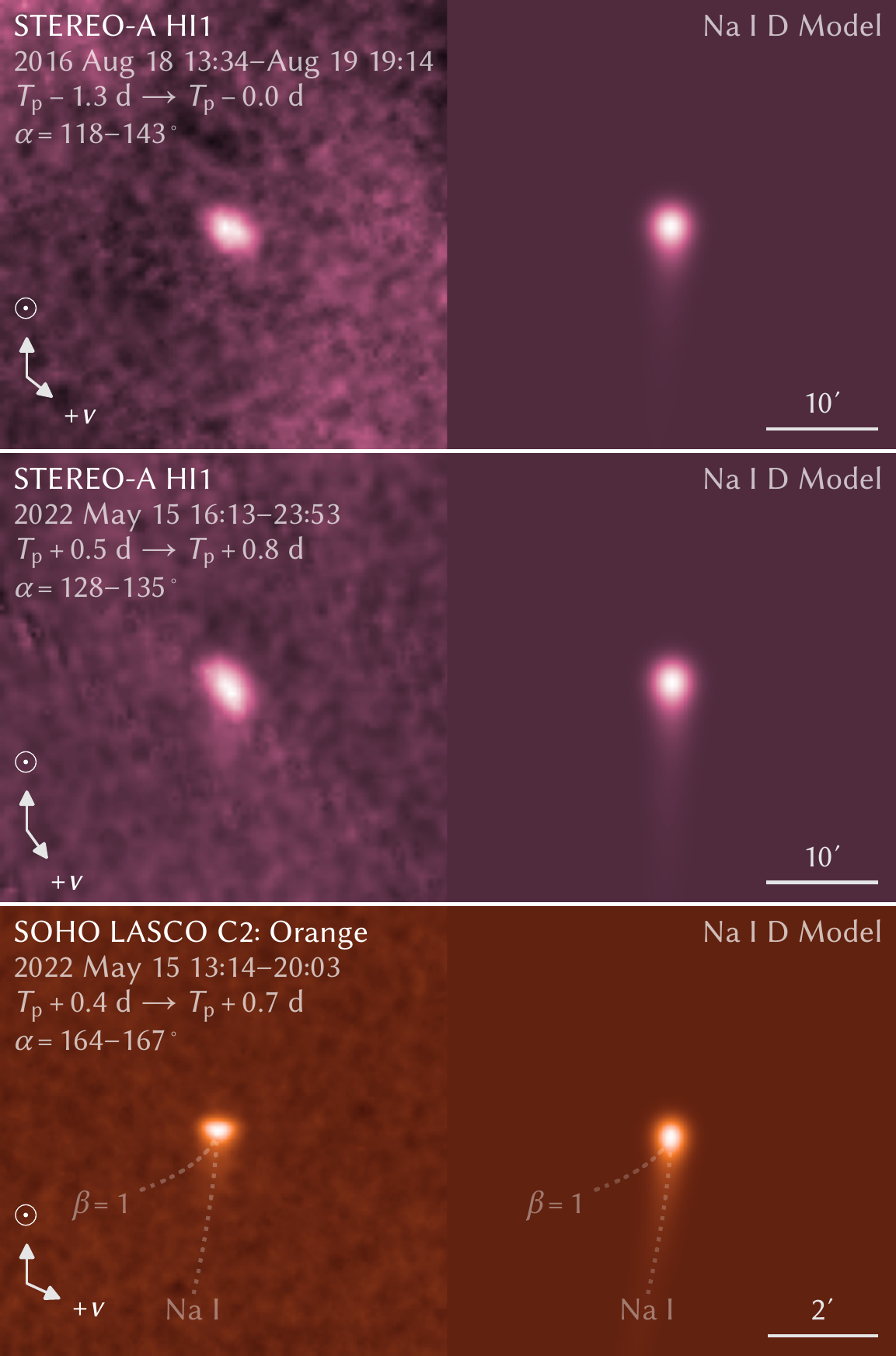}
\caption{\emph{Top:} Visualization of the modeled \ion{Na}{1} D tail brightness profile from 3~days before to 3~days after perihelion ($T_\mathrm{p}$), smoothed to a FWHM of 250,000~km to approximate the HI1 appearance at $\alpha\sim90^\circ$, and normalized to the brightness of the head to illustrate the increasing brightness of the tail relative to the head following perihelion. \emph{Lower left:} Evolution in the lengths of tail containing 50--90\% of the total \ion{Na}{1} D emission, and the effective fluorescence efficiency of the first 3000--100,000~km of the tail and that of the full tail---defined as the mean number of \ion{Na}{1} D photons emitted per atom while within that section of tail, or equivalently, the total \ion{Na}{1} D photon emission rate from that tail section divided by the \ion{Na}{1} production rate. \emph{Lower right:} Comparison of Phaethon's observed and modeled \ion{Na}{1} D morphology in HI1 under similar observing geometry and stacked to similar S/N in 2016 before perihelion and in 2022 after perihelion, as well as in LASCO C2 at higher $\alpha$ and resolution in 2022 after perihelion. Models use symmetric $3'$ (HI1) and $30''$ (C2) FWHM PSFs to approximate the actual, slightly trailed PSFs. Dotted curves in the C2 panel trace the expected positions of a tail of $\beta=1$ (micron-sized) dust and one of \ion{Na}{1}, demonstrating that the observed tail is compatible only with the latter.}
\label{fig:tail}
\end{figure*}

The fluorescence rate of each \ion{Na}{1} atom varies as a function of not only heliocentric distance $r$, but also radial velocity $\dot{r}$ under the Swings effect \citep{swings1941} due to the influence of deep \ion{Na}{1} D Fraunhofer absorption lines in the solar spectrum, which drives resonance fluorescence. The actual tail brightness is further strongly modified by the Greenstein effect \citep{greenstein1958} of the significant variation in $r$ and $\dot{r}$ of \ion{Na}{1} along the tail. In Appendix~\ref{sec:natail}, we detail a model to compute the brightness profiles of predominantly optically thin \ion{Na}{1} tails encompassing these effects together with \ion{Na}{1} photoionization as functions of asteroid $r$ and $\dot{r}$. In this section, we apply this model to Phaethon to demonstrate that its observed morphology and photometry are both consistent with \ion{Na}{1} activity.

\subsection{Morphology}
\label{sec:morph}

One distinctive aspect of Phaethon's activity noticed by \citet{jewitt2013} and \citet{hui2017} is its antisunward tail, which seems to lengthen over the course of $\sim$1~day after perihelion, which both studies interpreted as micron-sized dust taking $\sim$1~day to be accelerated by radiation pressure into a tail. Part of this arises from Phaethon's activity being overall much brighter post-perihelion, and a brightening tail rising above the noise level could appear to be lengthening. However, the \ion{Na}{1} tail model expects a physical lengthening of the tail over this period as a consequence of the Greenstein effect.

\ion{Na}{1} atoms ejected from Phaethon prior to perihelion, when Phaethon itself has $\dot{r}<0$, will initially largely also have $\dot{r}<0$, but are accelerated by radiation pressure toward $\dot{r}\sim0$. At $\dot{r}=0$, the \ion{Na}{1} D lines coincide with the \ion{Na}{1} D absorption lines in the solar spectrum driving the fluorescence, which reduces the excitation rate and thus the acceleration of atoms. The lowered acceleration then slows their escape to $\dot{r}>0$, trapping them in this weakly fluorescing state for a prolonged period, suppressing the tail. In contrast, \ion{Na}{1} atoms ejected after perihelion, when Phaethon has $\dot{r}<0$, will largely have initial $\dot{r}>0$ which the positive acceleration only further increases, resulting in a bright tail of efficiently fluorescing \ion{Na}{1} rapidly accelerating antisunward. Mercury's \ion{Na}{1} tail behaves similarly by this mechanism, appearing considerably brighter after perihelion \citep{schmidt2010a}.

Figure~\ref{fig:tail} illustrates the modeled lengthening of Phaethon's tail over the days surrounding perihelion. It also demonstrates from stacking pre- and post-perihelion data from similarly high phase angles $\alpha$---where Phaethon's activity outshines its surface, as discussed in Section~\ref{sec:photometry}---to similar signal-to-noise ratios (S/N) that the brightness of the tail relative to the head is indeed higher post-perihelion than pre-perihelion.

We also performed syndyne analysis of Phaethon's tail in the 2022 C2 Orange stack, taking advantage of its high $\alpha\approx166^\circ$ that exaggerates the small deviations in tail direction from antisunward indistinguishable by earlier analyses of HI1 observations at lower $\alpha$ \citep{jewitt2013,hui2017}. Due to Phaethon's orbit around the Sun, the otherwise convenient, sunward-aligned frame is actually a rotating reference frame, so \ion{Na}{1} or dust initially accelerating antisunward will follow curved trajectories in this frame due to Coriolis acceleration, with slower particles exhibiting greater curvature. \citet{finson1968} defines a parameter $\beta$ as the ratio between the radiation pressure acceleration and solar gravity, which is nearly constant for dust grains of a given size. As sunlight is predominantly comprised of photons with a wavelength on the order of $\sim$1~$\mu$m, micron-sized grains have the highest scattering efficiency and thus the highest $\beta\sim1$ \citep{gustafson2001,kimura2017} corresponding to the least curved dust trajectories. \ion{Na}{1} atoms, however, have much higher $\beta\sim7$--75, depending on $\dot{r}$, so will form a tail that is much less curved than any dust tail. As Figure~\ref{fig:tail} shows, the curvature of the observed tail appears consistent with \ion{Na}{1}, but excludes $\beta=1$ dust.

Note that this tail morphology also excludes atomic oxygen (\ion{O}{1})---which is both abundant in meteoritic material \citep{lodders2021} and often observed as a dissociation product of cometary volatiles \citep{decock2013}---from responsibility for the observed flux. While its forbidden 557.7~nm, 630.0, and 636.4~nm [\ion{O}{1}] lines do fall within the Orange and outside the Blue filter bandpasses like \ion{Na}{1} D emission, \ion{O}{1} cannot form a tail resolvable by our data: These weak [\ion{O}{1}] lines support minimal momentum transfer from sunlight corresponding to only $\beta\sim4\times10^{-5}$ \citep{fulle2007}, sufficient to propel the atoms antisunward by just $\sim$10--40~km, or $\sim$0.001~px in the 2022 LASCO C2 frames, over their $\sim$0.5--1~d photoionization lifetime at $r\approx0.14$~au \citep{fulle2007,huebner2015}. We are furthermore unaware of any plausible, unseen parent compounds with $\beta\gg1$ that could have distributed \ion{O}{1} along the observed tail prior to dissociation, or any alternative candidate species that could plausibly match both the tail morphology and Blue--Orange contrast expected for \ion{Na}{1}.

\subsection{Photometry}
\label{sec:photometry}

The other major distinctive aspect of Phaethon's activity is its asymmetric light curve that consistently rises sharply at perihelion into a peak $\sim$0.5~days later, before fading and vanishing days later. To explore this behavior, we constructed light curves by first dividing the data into bins spanning a certain time interval, and performing a median stack of each bin to improve resistance to cosmic rays and other artifacts. For our primary science data, we used bin sizes varying from 45~min for the 2022 C2 Orange data, which had the highest S/N, to 6~h for C3 Clear data, with the lowest S/N. We measured photometry from each stacked frame within apertures of radii $\rho=45''$ in C2, $\rho=2'$ in C3, and $\rho=3'$ in HI1, which were selected to maximize S/N to point sources while ensuring robustness to typical errors in the astrometric solutions of each camera. These apertures were accompanied by background annuli with $2\rho$/$3\rho$ inner/outer radii, which were sufficiently large to have minimal tail contamination. We then convolved each frame with the photometric aperture and used the standard deviation within the background aperture as the measurement uncertainty.

\begin{figure*}
\centering
\includegraphics[width=0.47\linewidth]{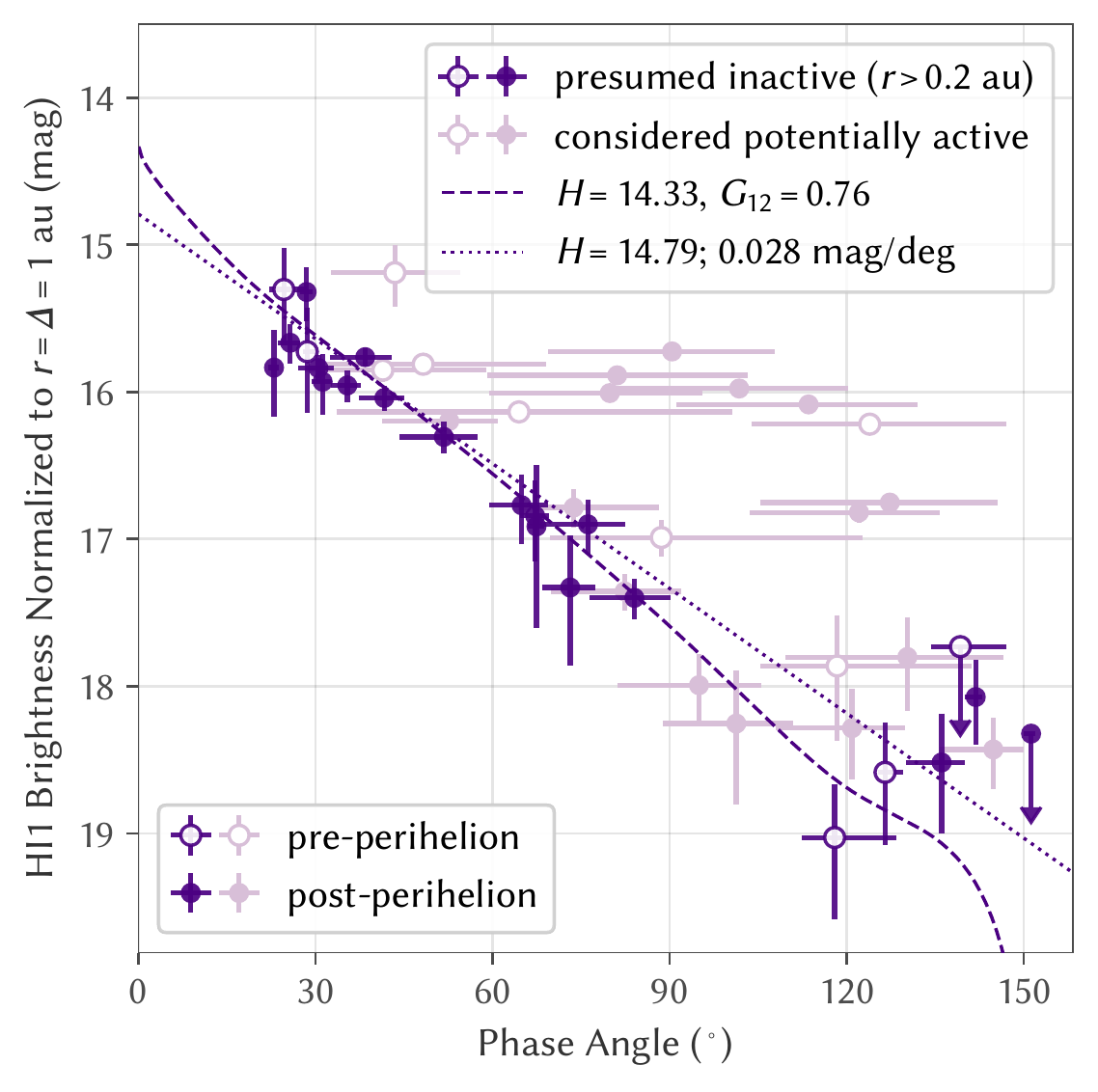}
\includegraphics[width=0.52\linewidth]{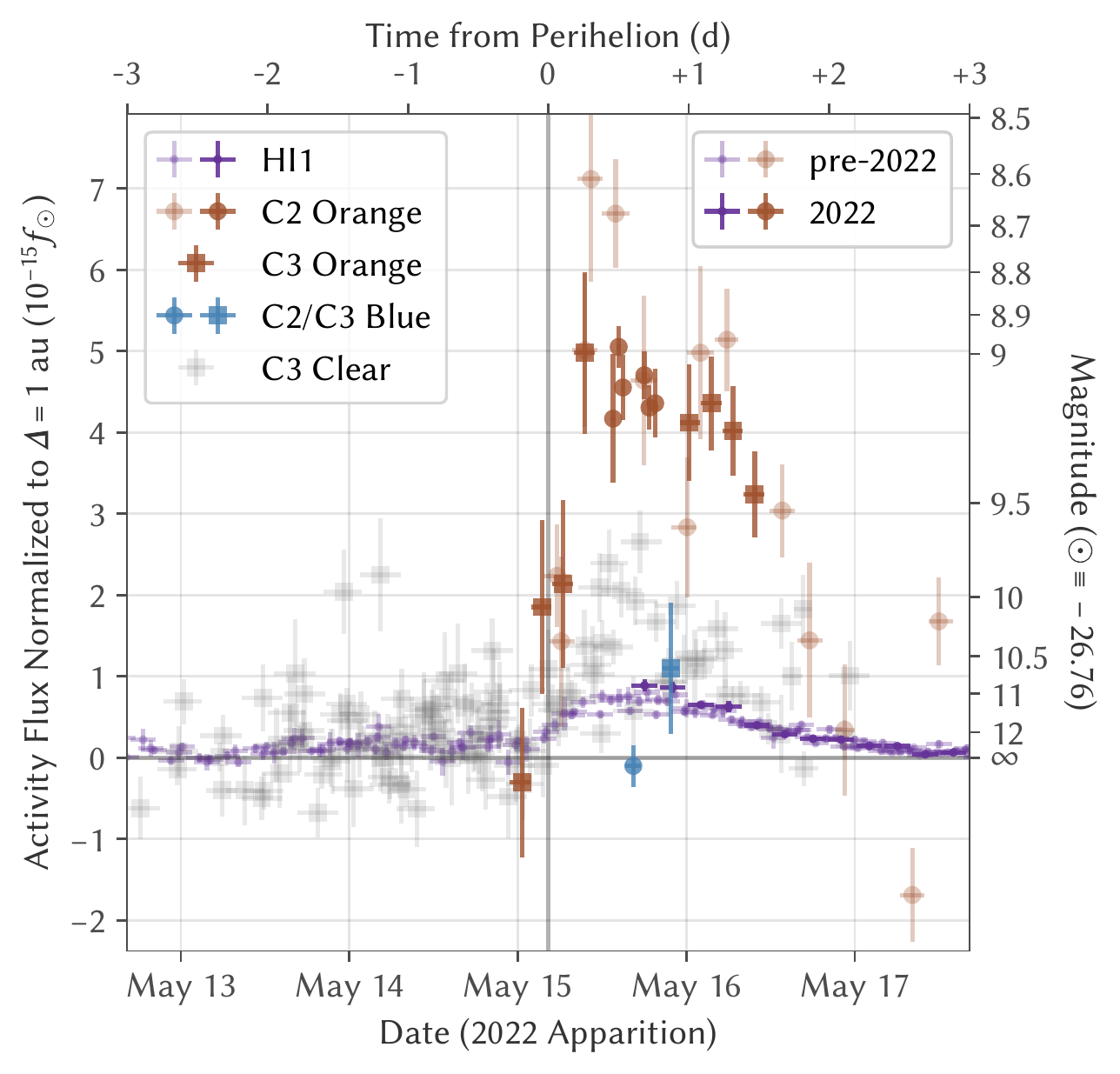}
\caption{\emph{Left:} Phase angle dependence of Phaethon's $r$- and $\varDelta$-normalized HI1 brightness. Excluding all bins extending to $r<0.2$~au isolates measurements presumed to be free of activity to yield our $H=14.33\pm0.10$, $G_{12}=0.76\pm0.29$ surface reflection model, with a linear fit shown for comparison. \emph{Right:} Surface-subtracted flux from all 18 apparitions, normalized to $\varDelta=1$~au, expressed relative to the mean solar flux through each bandpass ($f_\odot$). Dark/light symbols for C2 Orange and HI1 highlight observations from 2022/earlier apparitions. Note that these fluxes are measured within differently sized apertures of $45''$ radius for C2, $2'$ for C3, and $3'$ for HI1, so the plotted flux differences correspond only approximately to physical colors. Vertical error bars indicate $\pm1\sigma$, upper limits indicate $+3\sigma$, and horizontal error bars indicate the full range of the observations contributing to each point for both of these and all following plots.}
\label{fig:lc}
\end{figure*}

Before analyzing the activity, the flux from Phaethon's surface must be subtracted from the photometry. We measured this brightness directly by performing photometry of a separate copy of the HI1 data with $\sim$2~day time bins, then fitted an $H=14.33\pm0.10$, $G_{12}=0.76\pm0.29$ model \citep{muinonen2010} through only the points at $r>0.2$~au well beyond where activity has previously been reported, which we presume to capture only the inactive solid surface (as validated retrospectively by the \ion{Na}{1} production model we fit in Section~\ref{sec:nafit}). Spectra show that Phaethon has a slightly blue optical color, with reflectance differing by ${\lesssim}10\%$ between HI1 and the C2 and C3 bandpasses used \citep{binzel2001,deleon2010}. We consider this difference inconsequential for our purposes, and used this same fitted $HG_{12}$ model to subtract the surface contribution from the rest of our photometry, and thus isolate the brightness of the activity alone. Additionally, while the $HG_{12}$ model may not properly capture the fading of the surface at high $\alpha\gtrsim100^\circ$, the minimal surface contribution to overall brightness at these $\alpha$ becomes dwarfed by the phase-independent activity brightness, so contributes minimal error to the measured activity brightness. Figure~\ref{fig:lc} shows the result, normalized for observer distance, which reaffirms the initial finding that Phaethon's activity appears much brighter through Orange than any other filter. 

\begin{figure}
\centering
\includegraphics[width=\linewidth]{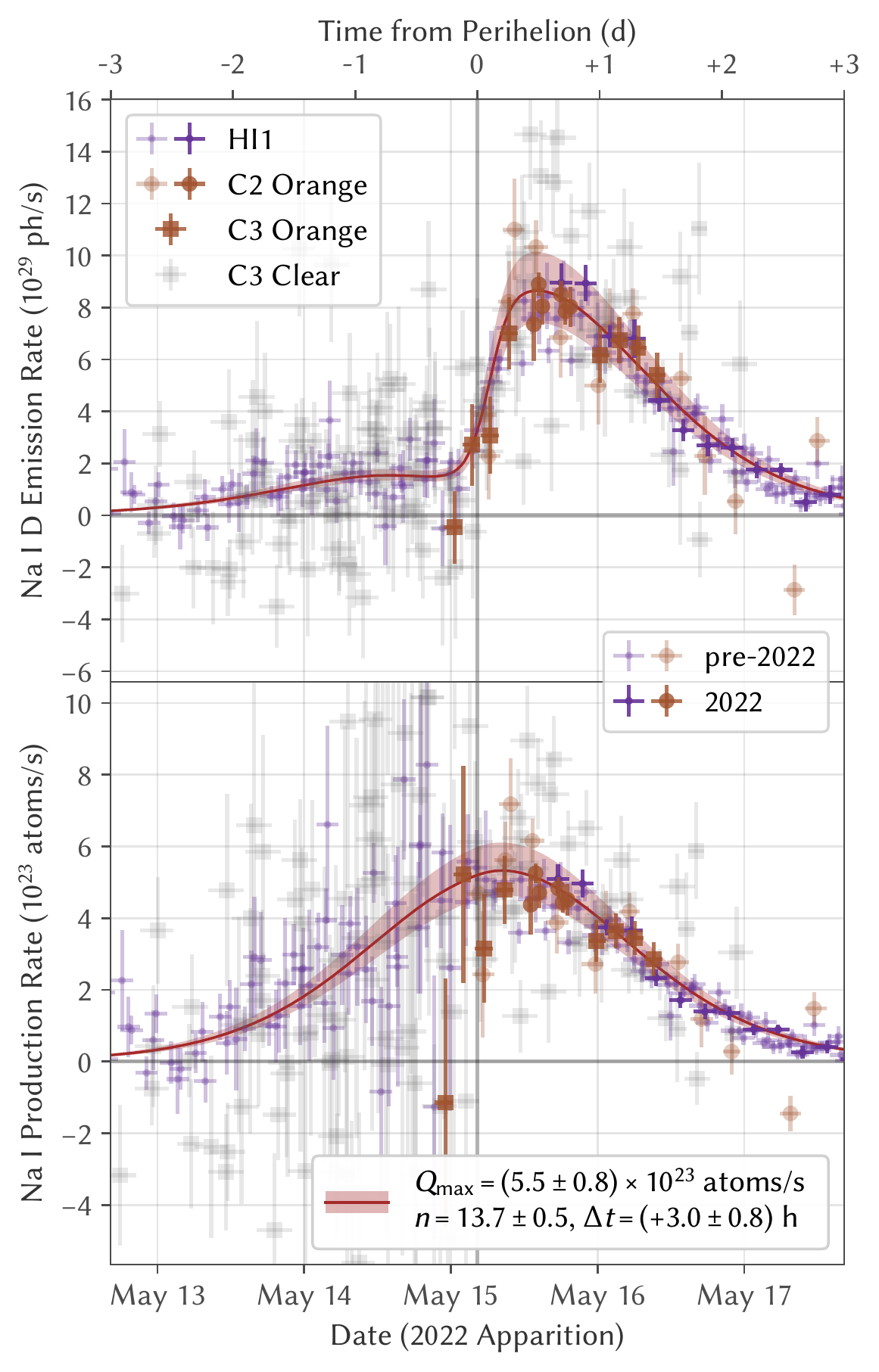}
\caption{Total \ion{Na}{1} D emission rate (top) and \ion{Na}{1} production rate (bottom) from Phaethon near perihelion as measured by LASCO and HI1 over all 18 observed apparitions, with the fitted model shown in brown. Darker symbols, again, highlight observations from the 2022 apparition. While fitted by the model, offsets between the different bandpasses have not been corrected for these plotted points to visually demonstrate the close agreement between the observed colors and those expected for \ion{Na}{1} D emission. We did, however, use the nominal, fitted $\tau_\mathrm{Na}=40$~h rather than the a priori 37~h to calibrate the plotted HI1 photometry, which amounted to a 13\% difference in those points.}
\label{fig:lc_na}
\end{figure}

Next, we used the \ion{Na}{1} tail model to translate the fluxes measured within the different photometric apertures to the \ion{Na}{1} emission rate of the full tail, as well as the corresponding \ion{Na}{1} production rate $Q(\text{\ion{Na}{1}})$. The results in Figure~\ref{fig:lc_na} show the Orange/Clear/HI1 emission and production rate curves to essentially coincide, demonstrating the associated colors remain consistent with those expected for \ion{Na}{1} D by the calibrations.

Moreover, unlike the light curves and \ion{Na}{1} D emission rates, the $Q(\text{\ion{Na}{1}})$ actually appears nearly symmetric about perihelion. The same Greenstein effect that suppresses the pre-perihelion length of the tail by the feedback effect described in Section~\ref{sec:morph} likewise suppresses the total brightness of the pre-perihelion tail. The pre-perihelion \ion{Na}{1} atoms spending a prolonged time with low fluorescence efficiency trapped at $\dot{r}\sim0$ will emit fewer \ion{Na}{1} D photons before photoionizing than post-perihelion atoms, which remain at $\dot{r}>0$ with high fluorescence efficiencies throughout their lifetime. The sudden brightness surge at perihelion therefore reflects not a surge in Phaethon's actual activity, but one in the overall fluorescence efficiency of Phaethon's \ion{Na}{1} tail, a quantity we define to be the total number of \ion{Na}{1} D photons emitted per atom released into the tail---equivalent to the \ion{Na}{1} D photon emission rate from the tail divided by $Q(\text{\ion{Na}{1}})$---as plotted in Figure~\ref{fig:tail}.

\begin{figure}
\centering
\includegraphics[width=\linewidth]{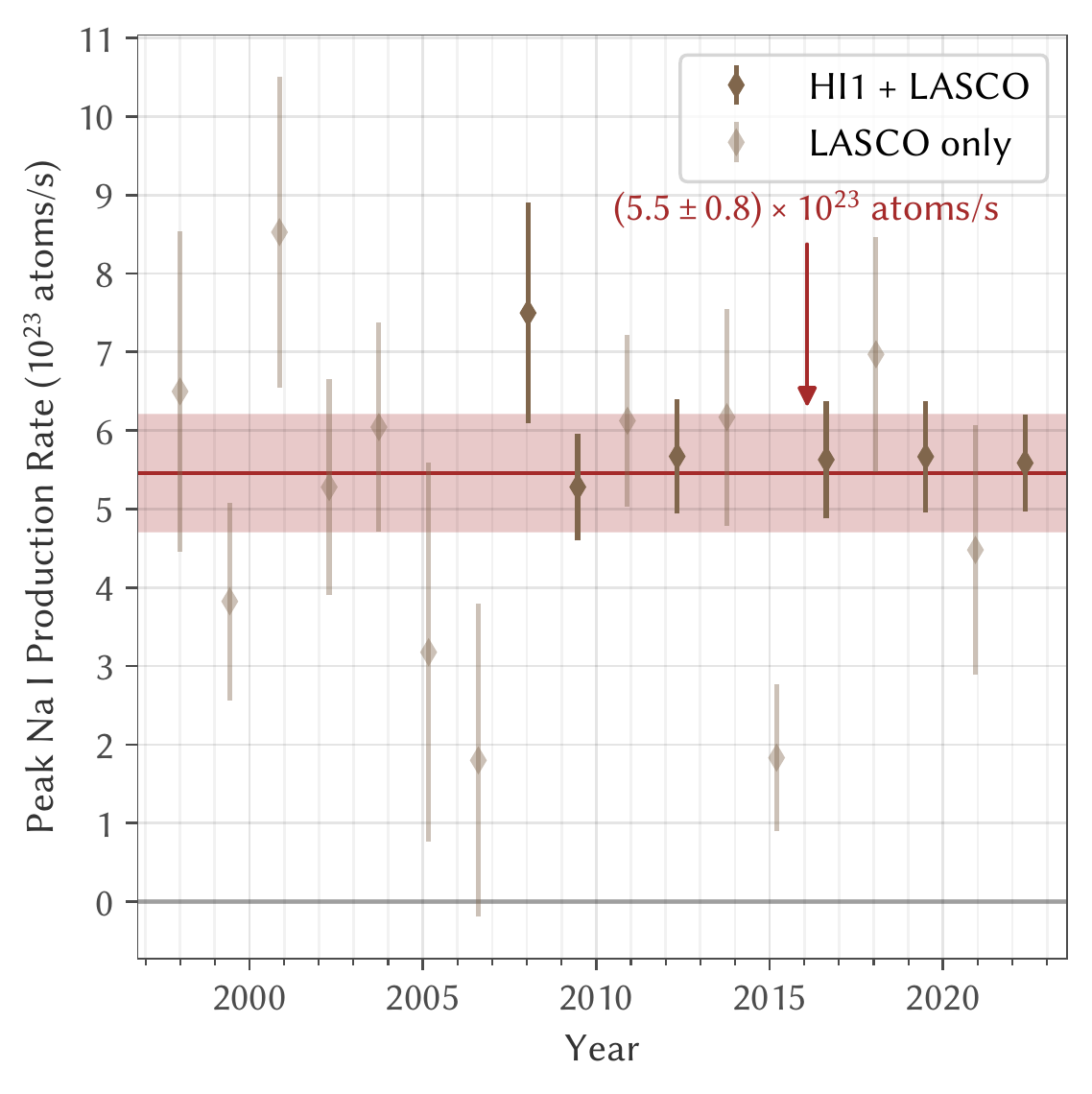}
\caption{Peak \ion{Na}{1} production rate $Q_\mathrm{max}$ at each apparition, fitted with all other parameters constrained to the results of the all-apparition fit. No trends appear evident, with the $Q_\mathrm{max}$ of individual apparitions generally consistent with the all-apparition fit of $(5.5\pm0.8)\times10^{23}$~atoms~s$^{-1}$.}
\label{fig:lc_peak}
\end{figure}

\subsubsection{Sodium Production Fit}
\label{sec:nafit}

To quantify Phaethon's \ion{Na}{1} production, we used a Markov-chain Monte Carlo (MCMC) process with our \ion{Na}{1} tail model to fit the photometry. We used a functional form

\begin{equation}
Q(\text{\ion{Na}{1}})=Q_\mathrm{max}\times\left(\frac{q}{r_{\Delta t}}\right)^n
\end{equation}

\noindent for the production rate $Q(\text{\ion{Na}{1}})$ at a given time, where $r_{\Delta t}$ is the $r$ of Phaethon at time $\Delta t$ earlier. The $q=0.140$~au is Phaethon's perihelion distance, while $Q_\mathrm{max}$, $n$, and $\Delta t$ are fittable, physical parameters, with $Q_\mathrm{max}$ being the peak $Q(\text{\ion{Na}{1}})$, $n$ indicating the $r^{-n}$ dependence of $Q(\text{\ion{Na}{1}})$, and $\Delta t$ being the offset of $Q(\text{\ion{Na}{1}})=Q_\mathrm{max}$ from perihelion. We used a log-uniform prior for $Q_\mathrm{max}$ and uniform priors for $n$ and $\Delta t$.

We also included several extra parameters to capture systematic effects from imperfect calibrations, modeling, and data reduction. First, we added the $r=1$~au \ion{Na}{1} lifetime $\tau_\mathrm{Na}$ (which scales to the actual \ion{Na}{1} lifetime at $r$ as $\tau_\mathrm{Na}\times r^2$) as a free parameter with a log-uniform prior to capture the otherwise uncharacterized uncertainty associated with our chosen $7.59\times10^{-6}$~s$^{-1}$ photoionization rate at $r=1$~au \citep{huebner2015}, which actually differs from earlier values by several tens of percent \citep[c.f.,][]{huebner1992,fulle2007}. We also use this value to correct the HI1 \ion{Na}{1} D sensitivity calibration from Appendix~\ref{sec:hi1sens}, which is sensitive to $\tau_\mathrm{Na}$ through its reliance on the brightness of distant portions of Mercury's \ion{Na}{1} tail.

We then introduced three parameters representing the offsets from the calibrated Orange, Orange--Clear difference, and HI1 magnitudes. Note that these offsets may not necessarily reflect only errors in the photometric calibrations, but could also capture errors in the modeled \ion{Na}{1} tail profiles since Orange (mostly C2), Clear (C3), and HI1 observations tend to capture flux from different lengths of tail due to photometric aperture and field of view differences between the cameras. We used a normally distributed prior with mean 0~mag and standard deviation 0.2~mag for all three parameters as a crude, initial estimate for the absolute uncertainties of the calibrations.

We also introduced a variability parameter $\xi$ prior to allow for underestimated uncertainties or variability between apparitions, and an outlier parameter $\epsilon$ to minimize the potential for large outliers to skew the fit, with flat priors constrained to positive values for both. We incorporated these parameters by modeling the residual likelihood distribution for each observation with calculated uncertainty $\sigma$ by a Voigt profile, the convolution of a normal distribution with standard deviation $\xi\sigma$ and a Cauchy--Lorentz distribution with scale parameter $\epsilon\sigma$.

We used emcee \citep{foreman-mackey2013} to sample the posterior distribution with 100 walkers. Following standard procedures, we considered the sampling converged at 50 times the maximum autocorrelation time estimated for the parameters, disposed of initial samples totaling two times the maximum autocorrelation time, and thinned the remaining samples by half the minimum autocorrelation time to obtain the posterior samples. We provide the resulting mean~$\pm~1\sigma$ values of all parameters from our fit to all 18 apparitions of Phaethon observations in Table~\ref{tab:nafit}, alongside those for equivalent fits to observations of 322P/SOHO and 323P/SOHO discussed in Section~\ref{sec:context}.

We overlaid the fitted $Q_\mathrm{max}=(5.5\pm0.8)\times10^{23}$~atoms~s$^{-1}$, $n=13.7\pm0.5$, and $\Delta t=(+3.0\pm0.8)$~h model over the data in Figure~\ref{fig:lc_na}, which demonstrates that it successfully reproduces the observed photometric behavior. Moreover, the fitted $\tau_\mathrm{Na}=(40\pm3)$~h (corresponding to an actual lifetime of $(0.78\pm0.06)$~h at $r=0.14$~au) is comparable to the $37$~h lifetime from the \citet{huebner2015} \ion{Na}{1} photoionization rate at $r=1$~au, while the fitted color offsets are all ${<}0.1$~mag, further reinforcing that the observed brightening arises from \ion{Na}{1} D emission. The $n=13.7\pm0.5$ is far steeper than that expected from \ion{Na}{1} production mechanisms like photon-stimulated desorption, solar wind ion sputtering, and meteoroid impact vaporization \citep{schmidt2012}, but is consistent with the sharp temperature dependence expected of thermal desorption which we therefore consider to be principally responsible for the observed \ion{Na}{1} activity. The small $\Delta t=(+3.0\pm0.8)$~h offset of peak \ion{Na}{1} production from perihelion is comparable to Phaethon's 3.6~h rotation period \citep{hanus2016}, and appears consistent with thermal lag for \ion{Na}{1} sourced from a depth on the order of $\sim$0.1~m, roughly the diurnal thermal skin depth \citep{masiero2021}.

We then explored the potential variability of Phaethon's activity between apparitions by repeating the MCMC sampling process for each individual apparition of Phaethon. We constrained these fits with a restrictive prior constructed from the all-apparition fitted posterior crudely approximated as a normal distribution with independent parameters, except leaving $Q_\mathrm{max}$ unconstrained with a uniform prior. The resulting $Q_\mathrm{max}$ of individual apparitions plotted in Figure~\ref{fig:lc_peak} show considerable scatter, but no clear trend. Note, in particular, the low scatter of apparitions with HI1 observations---especially after excluding 2008, where STEREO-B HI1 missed the post-perihelion peak---suggests the true variability in Phaethon's activity between apparitions is likely no more than a few percent over decade timescales.

\begin{deluxetable*}{lccccc}
\tablecaption{\ion{Na}{1} Model Parameters}
\label{tab:nafit}

\tablecolumns{6}
\tablehead{
\colhead{Parameter}  & \colhead{Units} & \colhead{(3200) Phaethon} & \colhead{322P/SOHO\tablenotemark{a}} & \colhead{323P/SOHO\tablenotemark{a}} & \colhead{Prior}
}

\startdata
perihelion distance\tablenotemark{b}, $q$ & au & 0.140 & 0.054 & 0.048 & fixed \\
\hline
\noalign{\vspace{0.8ex}}
\multicolumn{6}{c}{Fitted Parameters (Mean~$\pm~1\sigma$)} \\
\noalign{\vspace{0.8ex}}
\hline
peak \ion{Na}{1} production rate\tablenotemark{b}, $Q_\mathrm{max}$ & atoms~s$^{-1}$ & $(5.5\pm0.8)\times10^{23}$ & $(3.9\pm0.6)\times10^{25}$ & $(3.5\pm0.9)\times10^{24}$ & $Q_\mathrm{max}^{-1}$ \\
slope of \ion{Na}{1} production rate\tablenotemark{b}, $n$ & \nodata & $13.7\pm0.5$ & $8.8\pm0.3$ & $9.8\pm1.3$ & uniform \\
time offset of \ion{Na}{1} peak\tablenotemark{b}, $\Delta t$ & h & $+3.0\pm0.8$ & $-0.5\pm0.2$ & $-1.0\pm0.5$ & uniform \\
\ion{Na}{1} lifetime $\propto r^2$ at $r=1$~au, $\tau_\mathrm{Na}$ & h & $40\pm3$ & [$41\pm2$]\tablenotemark{c} & [$42\pm3$]\tablenotemark{c} & $\tau_\mathrm{Na}^{-1}$ \\
Orange photometry offset from calibration & mag & $+0.01\pm0.16$ & [$-0.03\pm0.10$]\tablenotemark{c} & [$-0.02\pm0.16$]\tablenotemark{c} & $0\pm0.2$\tablenotemark{d} \\
Clear--Orange offset from calibration & mag & $-0.09\pm0.06$ & [$-0.11\pm0.06$]\tablenotemark{c} & [$-0.08\pm0.06$]\tablenotemark{c} & $0\pm0.2$\tablenotemark{d} \\
HI1 photometry offset from calibration & mag & $-0.00\pm0.15$ & [$-0.27\pm0.12$]\tablenotemark{c} & \nodata & $0\pm0.2$\tablenotemark{d} \\
fractional photometric variability\tablenotemark{e}, $\xi$ & \nodata & $0.08\pm0.03$ & $0.36\pm0.03$ & $0.32\pm0.11$ & $\xi>0$ \\
relative outlier wing width\tablenotemark{e}, $\epsilon$ & \nodata & $0.10\pm0.04$ & $0.14\pm0.06$ & $0.20\pm0.11$ & $\epsilon>0$ \\
\hline
\noalign{\vspace{0.8ex}}
\multicolumn{6}{c}{Derived Properties (Mean~$\pm~1\sigma$)} \\
\noalign{\vspace{0.8ex}}
\hline
total \ion{Na}{1} production, $\bar{Q}$ & atoms~orbit$^{-1}$ & $(1.09\pm0.15)\times10^{29}$ & $(3.7\pm0.6)\times10^{30}$ & $(1.6\pm0.4)\times10^{29}$ & \nodata \\
Clear photometry offset from calibration & mag & $-0.08\pm0.16$ & \nodata & \nodata & \nodata \\
Orange--HI1 offset from calibration & mag & $+0.01\pm0.13$ & \nodata & \nodata & \nodata \\
Clear--HI1 offset from calibration & mag & $-0.08\pm0.13$ & \nodata & \nodata & \nodata
\enddata

\tablenotetext{a}{322P and 323P models may not accurately capture the physics of those comets as they exhibit systematic residuals to a much greater degree than the Phaethon model, so physical parameters likely have true errors far exceeding the computed uncertainties.}
\tablenotetext{b}{$Q(\text{\ion{Na}{1}})=Q_\mathrm{max}\times(q/r_{\Delta t})^n$, where $r_{\Delta t}$ is $r$ at time $\Delta t$ earlier.}
\tablenotetext{c}{Bracketed values were constrained with normally distributed priors with the mean~$\pm~1\sigma$ of the values fitted for Phaethon, so should not be used except as crude indicators of modeling error.}
\tablenotetext{d}{Normally distributed priors with listed mean~$\pm~1\sigma$.}
\tablenotetext{e}{Residual probability distribution for a point with formal uncertainty $\sigma$ is modeled as a core normal distribution with standard deviation $\xi\sigma$ convolved with an outlier Cauchy--Lorentz distribution with scale parameter $\epsilon\sigma$.}

\end{deluxetable*}
\vspace{-2.5em}

\section{Context}
\label{sec:context}

\begin{figure}
\centering
\includegraphics[width=0.99\linewidth]{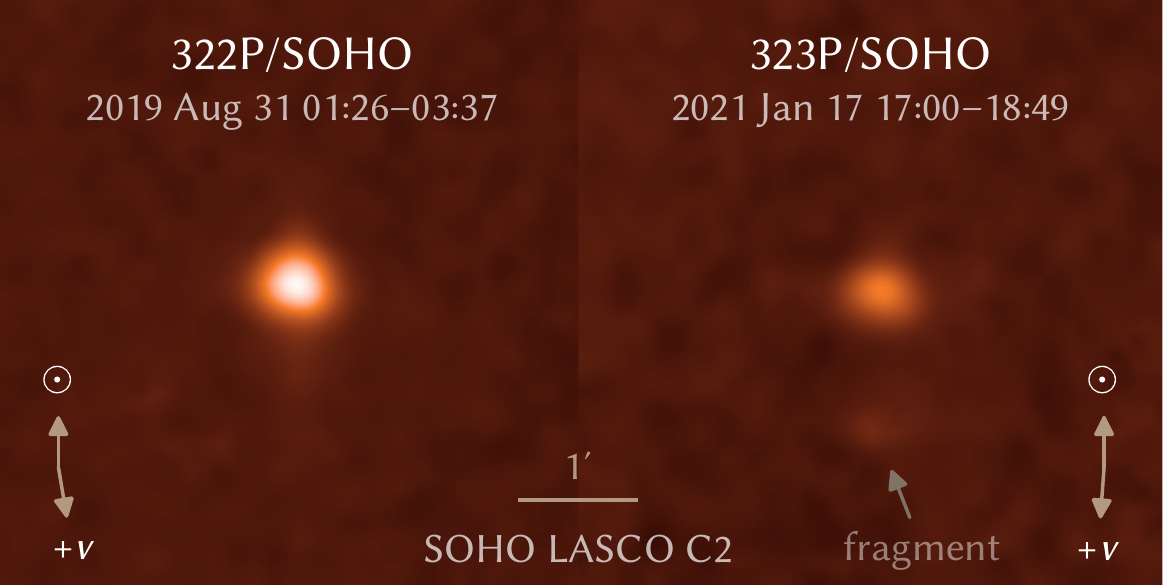}
\includegraphics[width=0.99\linewidth]{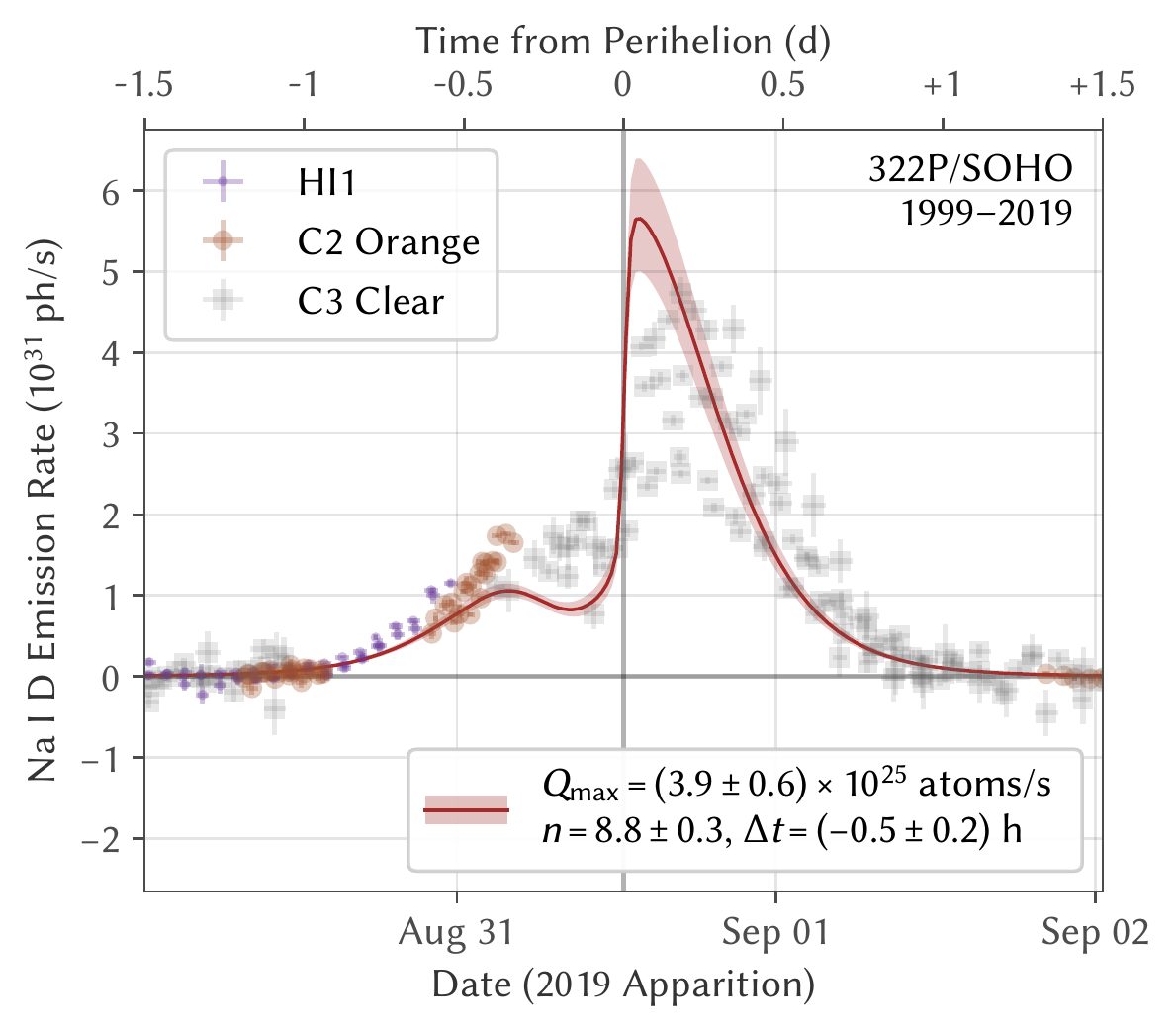}
\includegraphics[width=0.99\linewidth]{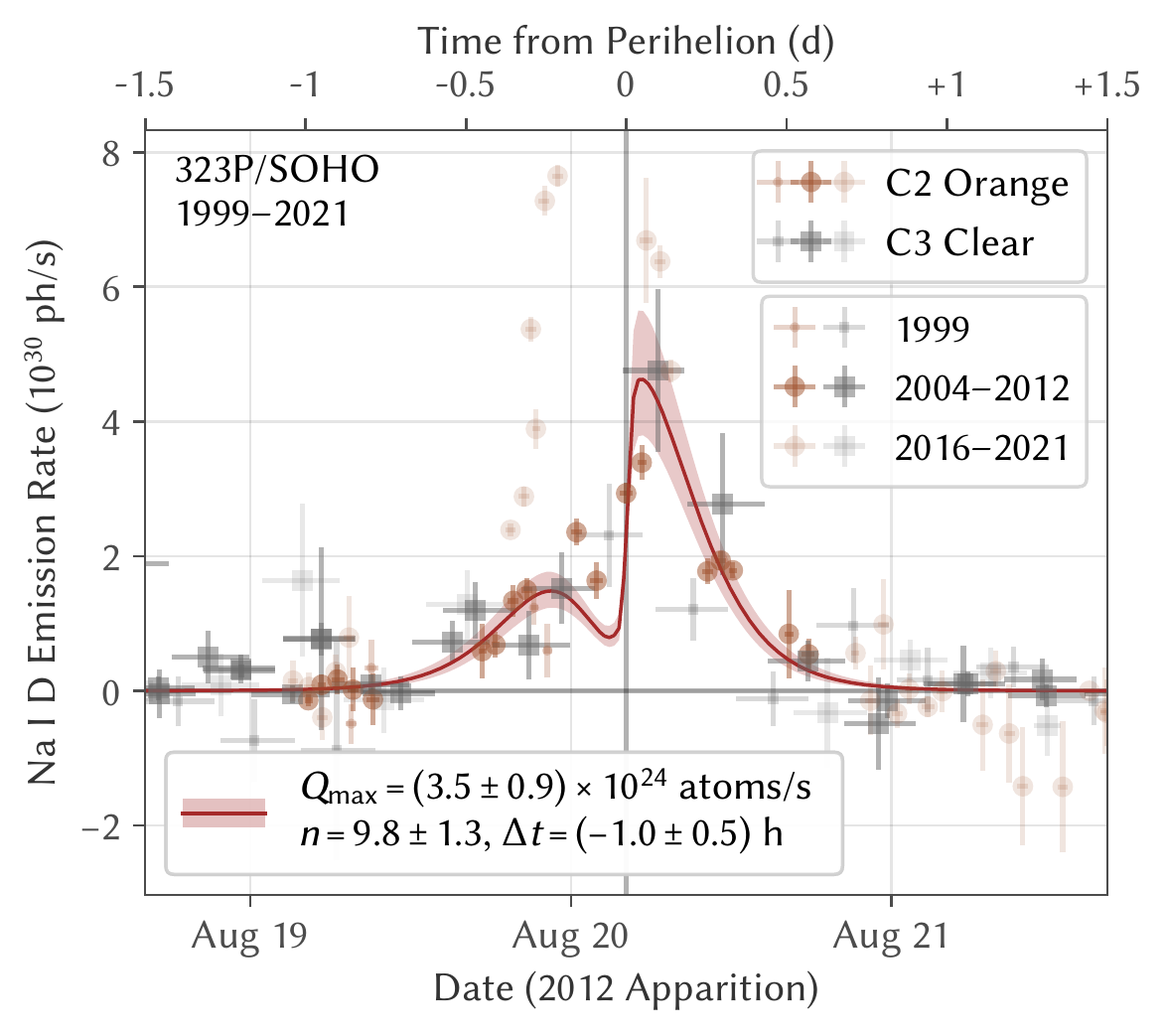}
\caption{\emph{Top:} 322P/SOHO and 323P/SOHO seen by LASCO C2 during their latest apparitions. Note the presence of a fragment leading 323P/SOHO by ${\sim}1'$. \emph{Middle/Bottom:} \ion{Na}{1} D emission rates for 322P/323P, with models fitted to all apparitions of 322P, and to the 2004--2012 apparitions of 323P while its orbit had $q=0.048$~au. These models do not fit the observations as well as the one for Phaethon, possibly due to some combination of spatially extended \ion{Na}{1} production and optical depths higher than those valid for the approximations used.}
\label{fig:32XP}
\end{figure}

\begin{figure*}
\centering
\includegraphics[width=\linewidth]{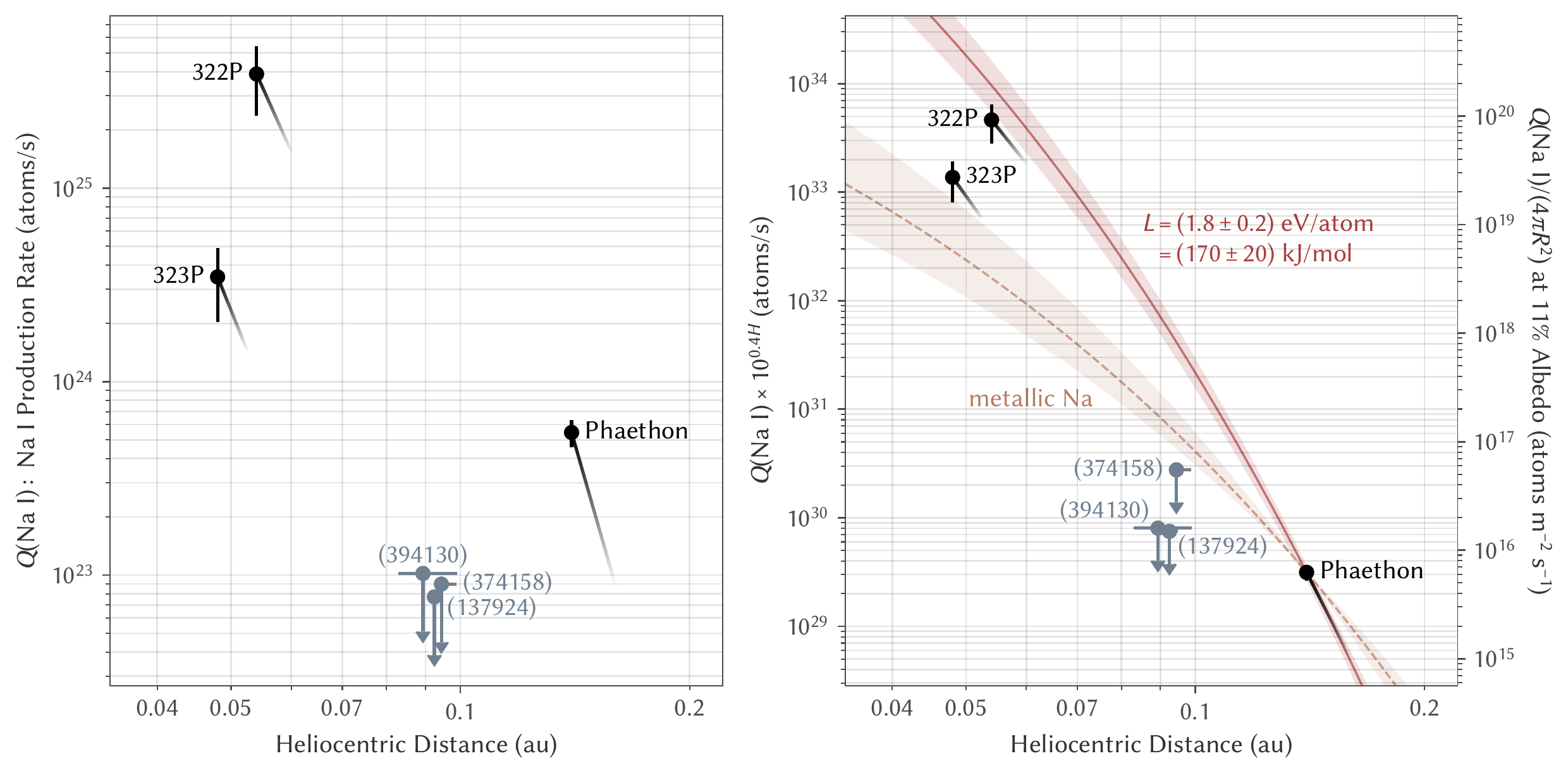}
\caption{\emph{Left:} \ion{Na}{1} production of Phaethon, 322P, and 323P at perihelion with mean slopes indicated by a faded line, as well as the $3\sigma$ upper limits for the asteroids (137924), (174158), and (394130). \emph{Right:} \ion{Na}{1} production multiplied by $10^{0.4H}$ (proportional to \ion{Na}{1} production per unit reflection cross section) for the same objects. The $n=13.7\pm0.5$ of Phaethon's activity is considerably steeper than expected for sublimation of metallic Na, and instead implies a modified latent heat of $L=(1.8\pm0.2)$~eV~atom$^{-1}=(170\pm20)$~kJ~mol$^{-1}$. The uncertainties illustrated for the metallic Na and modified $L$ extrapolations reflect uncertainties in Phaethon's \ion{Na}{1} production rate and subsurface temperature, as elaborated in Appendix~\ref{sec:desorp}.}
\label{fig:compare}
\end{figure*}

\subsection{Sodium Volatility on Sunskirting Objects}

\ion{Na}{1} D emission has routinely been documented in the spectra of nearly all bright comets observed at $r\lesssim0.6$~au since 1882 \citep[e.g.,][]{huggins1882,newall1910,adel1937,evans1967} and occasionally at even $r\sim1$~au \citep{oppenheimer1980,cremonese1997}. Analysis of the spatial and velocity profiles often show the \ion{Na}{1} to originate both from an extended, coma source---likely dust grains---and directly from the nucleus itself \citep{combi1997,brown1998,schmidt2015}.

In contrast, \ion{Na}{1} D emission has never previously been reported from any formally designated asteroid, like Phaethon, on which classical cometary volatiles are assumed to be absent. However, \citet{masiero2021} found that at the temperatures Phaethon experience at perihelion, the Na content of chondritic asteroidal material will volatilize and escape, which they propose could even drive comet-like dust production. With a fresh surface initially containing Na in metallic form at a roughly chondritic 0.5\% abundance by mass, they estimated Phaethon could attain $Q_\mathrm{max}\sim6\times10^{26}$~atoms~s$^{-1}$ near perihelion, or three orders of magnitude above the peak production rate we derive. This difference is not entirely surprising as Na will become increasingly depleted near the surface over repeated apparitions if not replenished, thus lowering the production rate over time. Most of the Na is also likely bonded within silicate materials rather than present in pure metallic form, which can change its thermal properties. The apparent depletion of Na in most Geminids meteoroids provides further evidence for such Na volatilization near Phaethon's orbit \citep{abe2020}.

\subsubsection{322P/SOHO and 323P/SOHO}

Many of SOHO's sunskirting comets may also be asteroids whose Na content has been volatilized by their proximity to the Sun. Two of these, 322P/SOHO and 323P/SOHO, have been recovered by nighttime telescopes as inactive nuclei at $r\sim1$~au where classical comets driven by water ice sublimation are active \citep{knight2016,hui2022}. Both nuclei also exhibited asteroid-like characteristics, being much smaller in size, displaying bluer colors, and at least 322P featuring a much higher albedo than typical cometary nuclei. Like Phaethon, these comets exhibit a strongly orange photometric color while active with little phase angle dependence, indicative of \ion{Na}{1} D emission with a lack of the micron-sized dust grains typically characteristic of active comets, but approach the Sun to a much closer $q\sim0.05$~au than Phaethon \citep{llebaria2006,knight2010}.

LASCO has observed six apparitions each of 322P and 323P since 1999. We reduced the photometry of all apparitions in the same manner as for Phaethon, and plot the resulting \ion{Na}{1} D emission rates in Figure~\ref{fig:32XP}. Both 322P and 323P appear much brighter than Phaethon, by $\sim$4~mag and $\sim$2~mag, respectively. They also vary in brightness between apparitions to a much greater degree, with 322P fluctuating by a factor of two and 323P by several times. Much of the variability of 323P arises from its highly erratic orbit, which brought its perihelion distance from $q=0.052$~au in 1999 to $q=0.048$~au for 2004, 2008, and 2012, then down to $q=0.039$~au for 2016 and 2021; isolating the 2004--2012 apparitions reduces the variability to the photometric noise level.

We then fit this data with the same MCMC procedure as described for Phaethon in Section~\ref{sec:nafit}, except with $\tau_\mathrm{Na}$ and the color offsets constrained by priors of normal distributions with the mean~$\pm~1\sigma$ of the values fitted for Phaethon. Results are again presented in Table~\ref{tab:nafit}. Note, however, that neither comet's light curve appears as well fit by this \ion{Na}{1} model: While they share the qualitative similarity of a pre-perihelion plateau in brightness followed by a sharp increase to a post-perihelion peak, the model overestimates the sharpness of the peak as well as the degree of pre/post-perihelion asymmetry for both comets. Therefore, one or more of the model assumptions must break down for these comets and the fitted parameters, provided in Table~\ref{tab:nafit}, should be treated as only rough estimates. Likely sources of error include significant contributions to \ion{Na}{1} production from extended sources like an unseen dust tail as well as optical depth modeling limitations at the much higher $Q(\text{\ion{Na}{1}})$ of these comets.

\subsubsection{Less Active Sunskirting Asteroids}

A number of formally designated asteroids approach the Sun more closely than Phaethon, yet none have ever been seen to be active \citep{holt2022}. All are much smaller than Phaethon, but several are still sufficiently large for observable activity if they had surfaces of comparable volatility. We selected favorable apparitions for a trio of asteroids with $q<0.1$~au: (137924) 2000 BD$_{19}$ in 2018, (374158) 2004 UL in 2009, and (394130) 2006 HY$_{51}$ in 2014. In each case, the asteroid crossed the LASCO C2 field of view at high phase angles immediately after perihelion, when \ion{Na}{1} tail fluorescence was most efficient. We processed all C2 Orange frames in the same manner as done for Phaethon, but were unable to detect any of the asteroids in the combined stacks.

Table~\ref{tab:compare} provides the $3\sigma$ upper limits for the asteroid trio---all ${\sim}10^{23}$~atoms~s$^{-1}$---in comparison to the fitted \ion{Na}{1} production rates for Phaethon, 322P, and 323P. We then approximately normalize the production rates for surface area by dividing by $10^{-0.4H}$, a quantity proportional to the geometric albedo times the cross sectional area, where the geometric albedo of all six objects likely all fall within a few factors of Phaethon's 11\% \citep{maclennan2022}. The results plotted in Figure~\ref{fig:compare} illustrate that the $3\sigma$ upper limits on normalized \ion{Na}{1} production by the asteroid trio cannot exclude a value comparable to that measured for Phaethon at its perihelion. However, these limits constrain the \ion{Na}{1} production of the asteroid trio closer to the Sun than Phaethon ever reaches, and Phaethon's extremely steep $n=13.7\pm0.5$ activity fall-off suggests its \ion{Na}{1} production would likely be much higher at those distances as well, complicating the comparison.

\begin{deluxetable*}{llllr}
\tablecaption{\ion{Na}{1} Production by Sunskirting Asteroids \& Comets}
\label{tab:compare}

\tablecolumns{5}
\tablehead{
\colhead{Name} & \colhead{$H$} & \colhead{Observation Period} & \colhead{$r$} & \colhead{$Q(\text{\ion{Na}{1}})$} \\
\colhead{} & \colhead{(mag)\tablenotemark{a}} & \colhead{(UT)} & \colhead{(au)\tablenotemark{b}} & \colhead{(atoms~s$^{-1}$)\tablenotemark{c}}
}

\startdata
(3200) Phaethon & $14.33\pm0.06$\tablenotemark{d} & 1997--2022 & 0.140 & $(5.5\pm0.9)\times10^{23}$ \\
322P/SOHO & $20.19$\tablenotemark{e} & 1999--2019 & 0.054 & $(3.9\pm1.5)\times10^{25}$ \\
323P/SOHO & $21.49\pm0.08$\tablenotemark{f} & 2004--2012 & 0.048 & $(3.5\pm1.5)\times10^{24}$ \\
(137924) 2000 BD$_{19}$ & $17.47$\tablenotemark{g} & 2018 Jul 25--26 & 0.092--0.095 & ${<}8\times10^{22}$ \\
(374158) 2004 UL & $18.72$\tablenotemark{g} & 2009 Apr 12 & 0.093--0.099 & ${<}9\times10^{22}$ \\
(394130) 2006 HY$_{51}$ & $17.24$\tablenotemark{g} & 2014 Nov 22 & 0.083--0.099 & ${<}1.0\times10^{23}$
\enddata

\tablenotetext{a}{Absolute magnitude, transformed to $V$ assuming solar color.}
\tablenotetext{b}{Heliocentric distance corresponding to provided $Q(\text{\ion{Na}{1}})$; equal to perihelion distance $q$ for Phaethon/322P/323P.}
\tablenotetext{c}{\ion{Na}{1} production rate at $r$. Uncertainties for Phaethon/322P/323P aim to capture variability and systematic fitting error, and are computed as $(\sigma(Q_\mathrm{max})^2+(\xi Q_\mathrm{max})^2)^{1/2}$, where $\sigma(Q_\mathrm{max})$ is the $1\sigma$ fit uncertainty of $Q_\mathrm{max}$. Upper limits are $3\sigma$.}
\tablenotetext{d}{From $HG_{12}$ fit in Figure~\ref{fig:lc}.}
\tablenotetext{e}{From $HG_1G_2$ fit by \citet{knight2016}.}
\tablenotetext{f}{From $HG_{12}$ fit by \citet{hui2022}.}
\tablenotetext{g}{From $H$, $G=0.15$ fit by the Minor Planet Center.}
\end{deluxetable*}
\vspace{-2.5em}

Appendix~\ref{sec:desorp} presents a rudimentary thermal desorption model for an isothermal blackbody asteroid to extrapolate Phaethon's \ion{Na}{1} to lower $r$ for this comparison. The steep $n=13.7\pm0.5$ appears inconsistent with the sublimation of metallic Na, but is well-modeled if the Na sequestered beyond $\sim$0.1~m below the surface had a modified $L=(1.8\pm0.2)~\mathrm{eV~atom^{-1}}=(170\pm20)~\mathrm{kJ~mol}^{-1}$. This derived $L$ agrees well with the empirically determined $\sim$1.8~eV~atom$^{-1}$ average binding energy for desorption of Na bound to oxide surfaces \citep{madey1998}. This value also falls within the 100--400~kJ~mol$^{-1}$ range previously measured for Kreutz sungrazing comets \citep{sekanina2003}, which may share a similar mechanism for \ion{Na}{1} activity.

We overlay both the metallic Na and modified $L$ extrapolations of Phaethon's normalized \ion{Na}{1} production to lower $r$ in Figure~\ref{fig:compare}. 322P falls very near the modified $L$ extrapolation, especially considering the additional uncharacterized uncertainties arising from thermal model simplifications, albedo differences, and the systematic errors in 322P's \ion{Na}{1} production model. In contrast, 323P falls an order of magnitude below the modified $L$ extrapolation, while the $+3\sigma$ limits for the asteroid trio fall 1--2 orders of magnitude below. The latter limits fall below even the shallower metallic Na extrapolation.

This initial comparison appears to show 322P's surface to be comparably volatile to Phaethon's surface, 323P's to be modestly less so, while those of (137924), (374158), and (394130) have become more deeply devolatilized. We therefore conclude that Phaethon's \ion{Na}{1} activity near perihelion, while not unique, is still unusual as it is not broadly shared among the overall population of sunskirting asteroids, so proximity to the Sun alone cannot explain its presently high surface volatility compared to other asteroids with comparable or lower $q$.

\subsection{Sodium-Driven Mass Loss Potential}

In the \citet{whipple1951} gas drag model, dust grains can be ejected from the surface when the outward gas drag they experience there exceeds the gravity holding them down. Phaethon's current $Q(\text{\ion{Na}{1}})=(5.5\pm0.8)\times10^{23}$~atoms~s$^{-1}$ near perihelion produces an average gas flux of ${\sim}2\times10^{16}$~atoms~m$^{-2}$~s$^{-1}$ over the ${\sim}4\pi\times(5.4~\mathrm{km}/2)^2$ surface area. With Phaethon near equinox at perihelion \citep{masiero2021}, solar heating should be fairly well-distributed over most of its surface and theoretically can support nearly isotropic subsurface Na desorption with minimal day--night variation. In practice, surface variations may effect significant local variations in \ion{Na}{1} flux, and we crudely estimate the peak \ion{Na}{1} flux at ${\sim}10^{17}$~atoms~m$^{-2}$~s$^{-1}$.

With a thermal \ion{Na}{1} outflow speed of $\sim$1~km~s$^{-1}$, the peak drag force on a dust grain of radius $R_g$ with a drag coefficient on the order of unity is ${\sim}10^{-12}~\mathrm{N}\times(R_g/1~\mathrm{mm})^2$. Estimates for the true density of both Geminid meteoroids and Phaethon vary widely, but assuming a typical bulk density of $\sim$2.6~g~cm$^{-3}$ for dust grains \citep{borovivcka2010} and $\sim$1.6~g~cm$^{-3}$ for the bulk asteroid itself, the surface gravity of ${\sim}10^{-3}$~m~s$^{-2}$ exerts a force of ${\sim}10^{-7}~\mathrm{N}\times(R_g/1~\mathrm{mm})^3$ on the grains, exceeding the drag force for all except very small grains of $R_g\lesssim1$~$\mu$m. While we technically cannot observationally constrain the abundance of submicron, Rayleigh scattering grains which rapidly drop in mass-normalized scattering efficiency $\propto R_g$, we placed tight bounds on the presence of micron-sized grains in Section~\ref{sec:2022app}, so consider Phaethon's current \ion{Na}{1} activity as unlikely to be driving significant dust production---at least on its own.

Phaethon, however, is also rotating with a rapid $3.6$~h period near the critical limit, producing an equatorial centrifugal acceleration ${\sim}10^{-3}$~m~s$^{-2}$ that nearly entirely offsets the gravitational acceleration in those regions. \citet{nakano2020} proposed that Phaethon's rotation was recently even faster and above the critical limit, which would have allowed it to shed dust grains and even boulders with no size limit. If Phaethon's effective surface gravity remains below $\sim$1\% of its non-rotating value anywhere on the surface, the observed \ion{Na}{1} activity could then lift $R_g\sim0.1$~mm grains from those areas which our high phase angle observations do not usefully constrain. However, such activity might be accompanied by purely rotationally driven mass loss from areas with slightly centrifugal acceleration that then exceeds the gravity, and the much greater efficiency of this process would likely marginalize the contribution of \ion{Na}{1} gas drag to dust production.

\ion{Na}{1} production could also theoretically alter Phaethon's rotation to indirectly drive mass loss through rotational instability, analogous to how the sublimation of icy volatiles visibly torque and disrupt comet nuclei \citep[e.g.,][]{bodewits2018}. \citet{marshall2022} recently reported Phaethon's rotational period to be decreasing at 4~ms~yr$^{-1}$, or a rotational acceleration of $2.1\times10^{-5}$~deg~day$^{-2}$. However, \ion{Na}{1} appears to again contribute incidentally at best: Under the most favorable setup of \ion{Na}{1} coherently directed along the equator, the observed, orbitally averaged \ion{Na}{1} production $\bar{Q}=(1.09\pm0.15)\times10^{29}$~atoms~orbit$^{-1}$ would torque Phaethon by $\sim$200~N~m. Even then, this maximum torque would rotationally accelerate Phaethon---treated as a uniform sphere---by only ${\sim}2\times10^{-7}$~deg~day$^{-2}$. More conventional thermal torque amplified by Phaethon's high surface temperatures and temperature gradients at perihelion may be a more plausible culprit in this case \citep[e.g.,][]{vokrouhlicky2015}, but will require more detailed modeling to verify.

\subsection{Sodium as a Tracer for Mass Loss}
\label{sec:tracer}

Regardless of its actual contribution to driving further mass loss, \ion{Na}{1} activity may still serve as an useful indicator for mass loss from any mechanism. Phaethon's current \ion{Na}{1} production depletes the $\sim$0.5\% by mass of Na from ${\sim}10^6$~kg~orbit$^{-1}$ of chondritic material \citep{lodders2021}, representing the mass loss required per orbit to sustain the \ion{Na}{1} production in steady state. This amount far exceeds the ${\lesssim}10^3$~kg of micron-sized dust ejected near perihelion, as found in Section~\ref{sec:2022app}---ruling out such dust as a major source of \ion{Na}{1}---but remains well below the ${\gtrsim}10^{10}$~kg~orbit$^{-1}$ needed to sustain the Geminids stream \citep{jewitt2010}. The true dust production may be much lower or entirely absent, as Phaethon's \ion{Na}{1} activity is not necessarily in steady state.

In the absence of unseen dust production clearing away devolatilized material covering Phaethon's surface, the devolatilized layer will gradually deepen and increasingly suppress the \ion{Na}{1} production rate over time. Measuring the Na abundance near the surface by comparing the \ion{Na}{1} production with that expected for a fresh surface can therefore, at least in theory, provide an estimate for when the surface was last cleared. In Appendix~\ref{sec:desorp}, we use our simple Na desorption model with crudely estimated surface characteristics to find that Phaethon's current $\sim$0.1~m layer of devolatilized material suppresses its \ion{Na}{1} production by a factor of ${\sim}10^{-3}$, corresponding to a surface devolatilized over ${\lesssim}10^4$~yr. This current bound is not yet particularly useful at constraining mass loss, as dynamical simulations indicate Phaethon has likely only spent a cumulative $\sim$10~kyr within at least the last 100~kyr with $q\lesssim0.15$~au where solar heating at perihelion is comparable to at present \citep{maclennan2021}, so the visible \ion{Na}{1} activity could, in theory, reflect only the brevity of Phaethon's sunskirting history.

However, the Geminids meteoroid stream did presumably form from Phaethon within the bounded time frame $\sim$1--10~kyr ago \citep{ryabova1999}. Ejection of the stream's ${\sim}10^{13}$~kg mass \citep{blaauw2017}---a sizable fraction of Phaethon's own ${\sim}10^{14}$~kg presently---was almost certainly associated with significant resurfacing of Phaethon that replenished its surface Na content from previously buried material. With better characterization of the surface material, and appropriate thermophysical and dynamical models, the observed \ion{Na}{1} production can conceivable set more stringent timing constraints. In this way, \ion{Na}{1} activity could serve as an observationally convenient, long-lasting record of significant mass loss in an asteroid's history even long after the mass loss event itself.

In Appendix~\ref{sec:desorp}, we calculate that Phaethon's \ion{Na}{1} production depletes the equivalent of only a $\sim$1~$\mu$m layer of subsurface material in a single apparition, explaining the lack of a discernible secular decline in \ion{Na}{1} production over decade timescales over which the devolatilized layer deepens by only a minute fraction. The much stronger \ion{Na}{1} activity of 322P and 323P, however, would deplete a more considerable $\sim$1--10~cm of material per apparition. Sustaining this level of activity over the multiple observed apparitions likely requires the surfaces of both objects to be actively eroding at perihelion, presumably driven by their high \ion{Na}{1} production and rapid rotation rates \citep{knight2016,hui2022}. Indeed, while 322P has never been observed away from the Sun after perihelion to ascertain its perihelic dust production, \citet{hui2022} found that 323P generated a debris field during its perihelion passage bright enough to be observed at $r\sim1$~au. The active fragment we found in LASCO C2 data from the 2021 apparition (see Figure~\ref{fig:32XP}) further reinforces this finding.

At $q\sim0.05$~au, both objects fall within the \citet{granvik2016} limit for $<$1~km diameter asteroids to be thermally disrupted, so their ongoing erosion appears likely to end only with their destruction. Likewise, the $q=0.08$--0.10~au of (137924), (374158), and (394130) place them beyond this zone, and their lack of observed \ion{Na}{1} activity indicates an absence of significant mass loss in at least the last few centuries, assuming decay timescales longer than a few decades like that of Phaethon. Phaethon itself has remained at even higher $q>0.1$~au over the past 100~kyr, reaching a minimum $q=0.126$~au only recently, $\sim$2~kyr ago \citep{williams1993,maclennan2021}---evidently too distant for mass loss to sustain itself and escalate into total disruption as 322P and 323P appear to be doing, given Phaethon's ongoing existence with minimal mass loss. \citet{ye2019} suggested through meteoroid stream analysis that asteroids on Phaethon-like orbits may still undergo partial disruption by some thermally related mechanism at an average $\sim$2~kyr cadence, but ultimately survive unlike objects much closer in.

While tempting to ascribe Phaethon's dramatically higher mass loss required during Geminids formation to increased solar heating at a slightly closer $q=0.126$~au than the present $q=0.140$~au, that difference alone would not obviously alter the key, qualitative characteristics of the \ion{Na}{1} activity. Direct extrapolation of Phaethon's modern \ion{Na}{1} production rate to $r=0.126$~au yields a four-fold increase from its current perihelion rate to $Q(\text{\ion{Na}{1}})\sim2\times10^{24}$~atoms~s$^{-1}$, which leaves previous order of magnitude estimates derived from the modern \ion{Na}{1} production substantially unchanged.

However, if the devolatilized layer were also much thinner or entirely absent, \ion{Na}{1} production could have been up to six orders of magnitudes higher through two effects discussed in more detail in Appendix~\ref{sec:desorp}: 

\begin{enumerate}
\item Removal of the $\sim$0.1~m devolatilized layer serving as a diffusive barrier to \ion{Na}{1} from below, increasing \ion{Na}{1} production by up to a factor of ${\sim}10^3$.
\item Na-bearing material exposed on the surface can reach subsolar temperatures of $\sim$1000~K far above the $\sim$700~K peak subsurface temperatures, which could raise overall \ion{Na}{1} production by another factor of ${\sim}10^3$.
\end{enumerate}

Such an extremely high $Q(\text{\ion{Na}{1}})\sim10^{30}$~atoms~s$^{-1}$ could potentially lift even meter-sized boulders without any rotational aid, but would require highly efficient surface clearing to sustain, as it would devolatilize a ${\sim}10$~m layer of material, or ${\sim}10^{12}$~kg---approaching the ${\sim}10^{13}$~kg of the entire Geminids stream---in one single apparition. Incomplete clearing of devolatilized surface material---particularly since only a fraction of the surface near the subsolar point can be maximally active at any time---would realistically lower this rate by one or more orders of magnitude, even if Phaethon were initially fully resurfaced in fresh, subsurface material.

Any such hypothetical elevation of \ion{Na}{1} production, however, requires a resurfacing mechanism to initially clear the devolatilized layer over a sizable portion of the surface. A disruptive trigger event---for example, from rotational instability or a large impact exposing subsurface material---must necessarily have reset the surface volatility to initiate any Na-supported Geminids formation. The upcoming DESTINY$^+$ flyby mission aims to provide resolved imaging of Phaethon's surface which could yield clearer evidence for such phenomena \citep{arai2018,ozaki2022}.

\subsubsection{Visibility of the Geminids Formation Process}

\begin{figure}
\centering
\includegraphics[width=\linewidth]{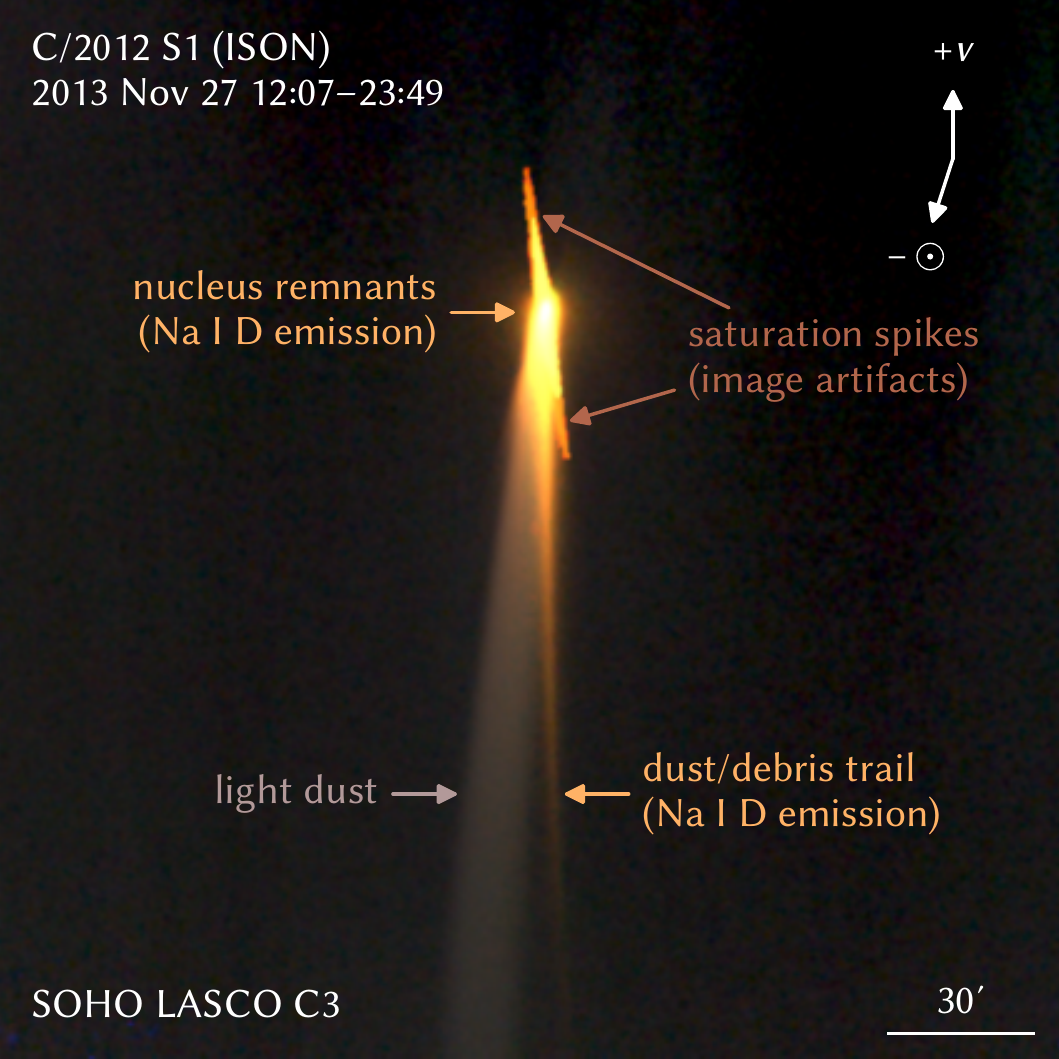}
\caption{LASCO C3 color composite (RGB = Orange/Clear/Blue) of the disintegrating Oort Cloud comet C/2012~S1 (ISON) at $r=0.1$~au, which may serve as a visual if not physical analog for a more active, Geminids-forming Phaethon. Bright \ion{Na}{1} D emission (golden color) highlights the remnants of the nucleus and a heavy ($\gtrsim$millimeter-sized) dust/debris trail. As with Phaethon's observed activity, this \ion{Na}{1} D emission far outshines the sunlight scattered by light (micron-sized) dust grains that typically dominates the optical brightness of active comets at similar $r$.}
\label{fig:2012S1}
\end{figure}

The recency of the Geminids formation within the past few thousand years raises the intriguing, if remote, prospect of the event being seen and recorded by ancient sky watchers during this period. A $Q(\text{\ion{Na}{1}})\sim10^{30}$~atoms~s$^{-1}$ is required for Phaethon to reach a magnitude of $V\sim-5$, necessary for naked eye observation in daylight at its ${\lesssim}7^\circ$ elongation at perihelion \citep{sekanina2022}. While pushing the theoretical upper limit for \ion{Na}{1} thermally desorbed directly from the surface, such a rate could be readily attained through the accompanying dust production which provides a means to far more efficiently excavate Na, since dust grains may still retain a sizable fraction of their Na until after ejection, thus dramatically increasing the total exposed surface area.

One potential modern analog to Phaethon during Geminids formation is the sungrazing comet C/2012~S1 (ISON), which brightened to $V\sim-3$ at $r\sim0.1$~au \citep{knight2014} roughly one week after its $\sim$1~km diameter nucleus \citep{lamy2014} apparently disintegrated \citep{sekanina2014}. As illustrated in Figure~\ref{fig:2012S1}, its intense brightness was confined almost exclusively to \ion{Na}{1} D emission, which corresponded to $Q(\text{\ion{Na}{1}})\sim10^{29}$~atoms~s$^{-1}$ that exceeded even \edit{1}{its own \edit{1}{$Q(\text{H}_2\text{O})=(2.2\pm0.5)\times10^{28}$~molecules~s$^{-1}$} at $r=1$~au \citep{combi2014}. This $Q(\text{\ion{Na}{1}})$ sustained over its $\sim$1~day period of near-peak activity produced a total of ${\sim}10^{34}$~atoms of \ion{Na}{1}, corresponding to the Na content in ${\sim}10^{11}$~kg of cometary material with a chondrite-like $\sim$0.5\% Na abundance by mass---comparable to the total mass of a $\sim$1~km diameter nucleus with a typical cometary bulk density of $\sim$0.6~g~cm$^{-3}$ \citep{weissman2008}. C/2012~S1's extreme but otherwise asteroid-like behavior therefore involves the loss of a sizable fraction if not a majority of its original Na content, likely facilitated by its earlier disintegration into a field of debris much smaller than the original nucleus. Such debris would have been rapidly depleted of icy cometary volatiles much farther from the Sun---eliminating the source of classical cometary activity---but would have retained most of its more refractory Na content until reaching sunskirting distances where this Na could be impulsively released.}

While C/2012~S1 was still too faint to be widely seen at its peak, the Geminids meteoroid stream has a combined mass on the order of ${\sim}10^{13}$~kg that is ${\sim}10^2\times$ larger \citep{blaauw2017}. The formation of the latter could therefore have involved considerably brighter events if a sizable fraction of the total mass was released in one or a few cataclysmic events. \edit{1}{An ejection of ${\sim}10^{12}$~kg of debris with a $\sim$0.5\% Na mass fraction contains ${\sim}10^{35}$~atoms of Na which, when released over the span of one to a few days near perihelion, provides the requisite ${\sim}10^{30}$~atoms~s$^{-1}$ for clear daylight visibility.}

The likely absence of a prominent, cometary tail of micron-sized dust \edit{1}{accompanying the bright \ion{Na}{1} D emission} would cause Phaethon and its debris to take a distinctly orange/red hue with nearly starlike naked eye morphology due to the short lifetime of \ion{Na}{1} at its perihelion. Although not directly related to the formation of the Geminids, this description does notably fit reports of a daylight object seen briefly in 1921 at Lick Observatory and a few other sites \citep{pearce1921,sekanina2016}, which may have been the \ion{Na}{1} D emission of debris from another disrupted asteroid or comet. Reports of similar starlike objects beside the Sun also appear in Chinese records, although the credibility of most of these claims cannot be reliably established \citep{strom2002}. A more thorough investigation into potential observations of Phaethon, or lack thereof, during its formation of the Geminids could provide a unique source of direct observational constraints on the process.

\section{Conclusions}

We observed the perihelic brightening of sunskirting asteroid and Geminids parent (3200) Phaethon with the SOHO LASCO coronagraphs and the STEREO HI1 imagers over a total of 18 apparitions. We used three distinct lines of evidence to demonstrate that this brightening is from \ion{Na}{1} D emission rather than dust:

\begin{enumerate}
\item Photometric colors: Phaethon's activity appears much brighter in Orange-filtered LASCO images than in Clear-filtered LASCO or HI1 images, and cannot be seen at all in Blue-filtered LASCO images. The measured colors match those expected for pure \ion{Na}{1} D emission by our calibrations.
\item Tail morphology: Phaethon's tail grows in length and intensity from before to after perihelion as expected from effects of radiation pressure, Doppler shift, and solar Fraunhofer lines for a tail of \ion{Na}{1}. The tail's curvature also appears consistent with \ion{Na}{1} and inconsistent with dust.
\item Light curve pattern: Phaethon's brightness is sharply asymmetric about perihelion, being much greater after than before perihelion, which is again well-modeled by a tail of \ion{Na}{1} with nearly symmetric \ion{Na}{1} production. The fitted \ion{Na}{1} lifetime of $(40\pm3)$~h at $r=1$~au is likewise consistent with its photoionization under solar radiation.
\end{enumerate}

We then analyzed the \ion{Na}{1} production of Phaethon, compared it to those of other sunskirting comets and asteroids, and drew several key conclusions and inferences:

\begin{enumerate}
\item Phaethon attains a consistent peak \ion{Na}{1} production rate of $(5.5\pm0.8)\times10^{23}$~atoms~s$^{-1}$ at $(3.0\pm0.9)$~h after perihelion, with a steep heliocentric distance dependence of $r^{-13.7\pm0.5}$.
\item The total $(1.09\pm0.15)\times10^{29}$~atoms~orbit$^{-1}$ corresponds to the depletion of Na from ${\sim}10^6$~kg~orbit$^{-1}$ of chondritic material, although Phaethon's actual ongoing mass loss may be much lower as \ion{Na}{1} activity is not necessarily sustained in steady state and could instead be decaying over timescales longer than a few decades.
\item Phaethon's \ion{Na}{1} activity is likely driven by thermal desorption of Na, which is bound with an effective latent heat of $(1.8\pm0.2)$~eV~atom$^{-1}=(170\pm20)$~kJ~mol$^{-1}$ and sequestered beneath an effective $\sim$0.1~m deep devolatilized layer.
\item Sunskirting comets 322P/SOHO and 323P/SOHO are likely asteroids experiencing high levels of heating at $r\sim0.05$~au sufficient to actively erode their surfaces, which clears away Na-depleted material to sustain their strong \ion{Na}{1} activity.
\item No activity was seen from sunskirting asteroids (137924), (374158), and (394130), indicating they exhibit much lower surface volatility than Phaethon, and likely have not experienced significant mass loss within the last few centuries. Sublimation-driven mass loss at their $r\sim0.1$~au, alone, does not appear capable of clearing Na-depleted material at a rate sufficient to sustain \ion{Na}{1} production indefinitely.
\item While \ion{Na}{1} gas drag could potentially drive further mass loss itself, \ion{Na}{1} production and thus \ion{Na}{1} D emission will accompany mass loss exposing fresh, subsurface material by any mechanism. Phaethon was likely much brighter from such emission during the Geminids formation period due to efficient excavation of subsurface Na, potentially even briefly to the point of daylight visibility.
\end{enumerate}

Future investigations extending our findings with more sophisticated thermophysical and dynamical modeling of Phaethon and the Geminids stream may provide more detailed insight into the mechanics of Phaethon's \ion{Na}{1} activity and its role, if any, in the still enigmatic Geminids formation process.

\bigskip % preprint
%\begin{acknowledgments} % submission
We thank Kevin Schenk (NASA/GSFC) and the SOHO mission operations team for carrying out our special observing sequence targeting Phaethon with LASCO, Chris Scott (Reading) for bringing the shift in the HI1 bandpass to our attention, Joe Masiero (Caltech/IPAC) for useful discussions on thermal desorption, and Gregg Hallinan (Caltech) for feedback and support. We also thank the anonymous reviewers for their helpful comments and suggestions toward improving this work.

K.~B. was supported by the NASA-funded Sungrazer Project. Q.~Y. was supported by STScI grant HST-GO-15357 and NASA program 80NSSC22K0772. M.~M.~K. was supported by NASA program 80HQTR20T0092. C.~S. acknowledges support from NASA programs 80NSSC19K0790 and 80NSSC22K1303. SOHO is a project of international cooperation between ESA and NASA.
%\end{acknowledgments} % submission

\facilities{SOHO (LASCO), STEREO (HI1)}

\software{Astropy \citep{astropy2022}, Astroquery \citep{ginsburg2019}, emcee \citep{foreman-mackey2013}, Matplotlib \citep{hunter2007}, NumPy \citep{vanderwalt2011}, sbpy \citep{mommert2019}, SciPy \citep{virtanen2020}}

\appendix

\section{HI1 Sodium Sensitivity}
\label{sec:hi1sens}

\begin{deluxetable*}{llccccc}
\tablecaption{Observations of Mercury's Tail Used for \ion{Na}{1} D Sensitivity Calibration}
\label{tab:mercobs}

\tablecolumns{7}
\tablehead{
\colhead{Instrument} & \colhead{Observation Time} & \colhead{$\alpha$} & \colhead{$\theta$} & \colhead{$Q(\text{\ion{Na}{1}})$} & \colhead{\ion{Na}{1} D Sensitivity} & \colhead{0~mag Flux} \\
\colhead{} & \colhead{(UT)} & \colhead{($^\circ$)\tablenotemark{a}} & \colhead{($^\circ$)\tablenotemark{b}} & \colhead{(atoms~s$^{-1}$)\tablenotemark{c}} & \colhead{(DN~ph$^{-1}$~m$^2$)\tablenotemark{d}} & \colhead{(ph~m$^{-2}$~s$^{-1}$)\tablenotemark{d,e}} 
}

\startdata
STEREO-A HI1 & 2007 Feb 20--21 & 158--166 & 62--70 & $1.25\times10^{24}$ & $(7.7\pm0.9)\times10^{-7}$ & $(4.2\pm0.5)\times10^{10}$ \\
STEREO-A HI1 & 2008 Feb 07--08 & 147--158 & 61--71 & $1.25\times10^{24}$ & $(6.4\pm0.7)\times10^{-7}$ & $(5.0\pm0.5)\times10^{10}$ \\
STEREO-B HI1 & 2010 Apr 09--10 & 148--159 & 62--73 & $1.25\times10^{24}$ &  $(7.3\pm0.8)\times10^{-7}$ & $(4.3\pm0.4)\times10^{10}$ \\
STEREO-A HI1 & 2011 Dec 27--28 & 158--166 & 112--119 & $6.3\times10^{23}$ &  $(6.6\pm0.7)\times10^{-7}$ & $(4.9\pm0.6)\times10^{10}$ \\
STEREO-A HI1 & 2016 Jul 14--15 & 155--164 & 68--76 & $1.24\times10^{24}$ &  $(8.5\pm1.0)\times10^{-7}$ & $(3.8\pm0.5)\times10^{10}$ \\
STEREO-A HI1 & 2020 May 30--Jun~02 & 152--167 & 105--120 & $7.1\times10^{23}$ &  $(5.6\pm0.6)\times10^{-7}$ & $(5.6\pm0.6)\times10^{10}$ \\
STEREO-A HI1 & 2021 May 20--21 & 159--167 & 116--123 & $5.3\times10^{23}$ &  $(6.2\pm0.7)\times10^{-7}$ & $(5.1\pm0.5)\times10^{10}$ 
\enddata

\tablenotetext{a}{Phase angle of Mercury from observing spacecraft.}
\tablenotetext{b}{True anomaly of Mercury.}
\tablenotetext{c}{Estimated \ion{Na}{1} escape rate with assumed $\pm10\%$ uncertainty, fitted from observations by \citet{schmidt2010a}.}
\tablenotetext{d}{For $r=1$~au equivalent \ion{Na}{1} ionization rate of $7.59\times10^{-6}$~s$^{-1}$ \citep{huebner2015}.}
\tablenotetext{e}{\ion{Na}{1} D flux producing magnitude 0 under the \citet{tappin2022} HI1A and \citet{tappin2017} HI1B calibrations.}
\end{deluxetable*}
\vspace{-2.5em}

\begin{figure*}
\centering
\raisebox{-\height}{\includegraphics[width=0.46\linewidth]{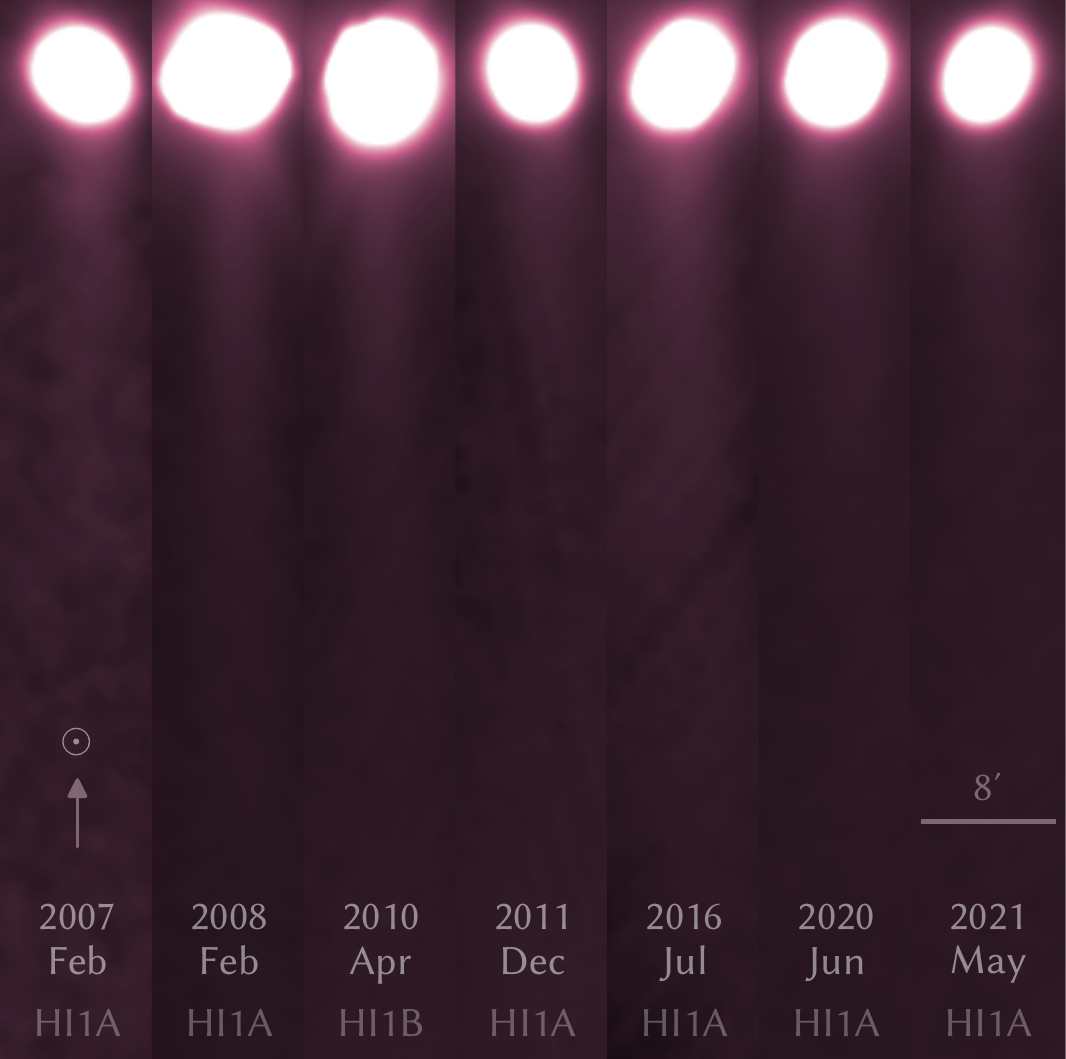}}
\raisebox{-\height}{\includegraphics[width=0.52\linewidth]{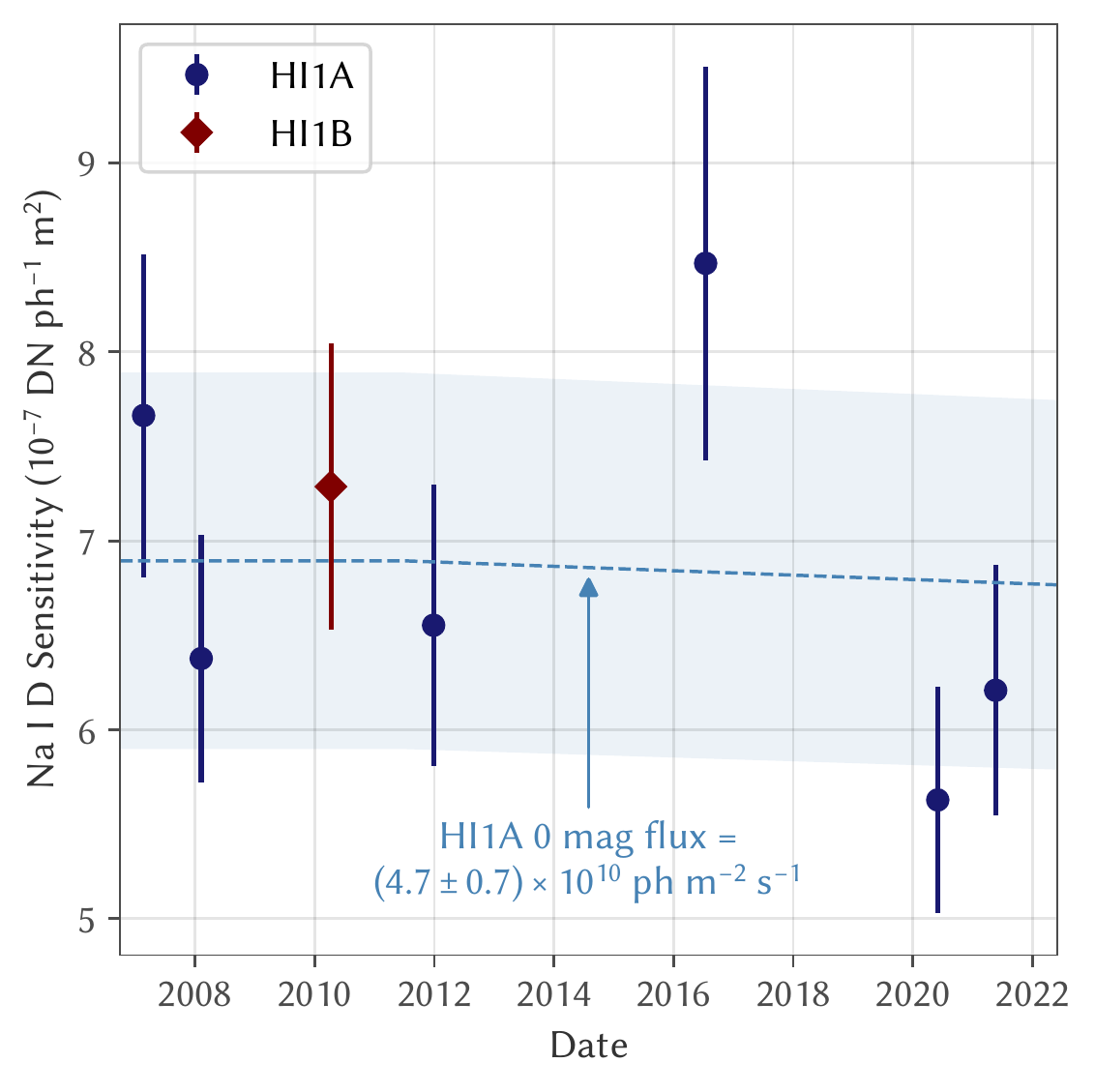}}
\caption{\emph{Left:} Mercury and its \ion{Na}{1} tail seen by the HI1 cameras of STEREO-A (HI1A) and STEREO-B (HI1B) at 7 epochs. The \ion{Na}{1} tail extends downward (antisunward) from Mercury in all frames; the diagonal branches extending from Mercury in the 2008 and 2010 frames are saturation artifacts, while diagonal stripes along the tail are residual star trails from imperfect background subtraction. \emph{Right:} HI1 sensitivity to \ion{Na}{1} D photons, as measured by the fitted brightness of Mercury's tail at each epoch. The computed 0~mag flux of $(4.7\pm0.7)\times10^{10}$~ph~m$^{-2}$~s$^{-1}$ (equivalent to a $V=0$ \ion{Na}{1} D source magnitude of $+0.89\pm0.16$) under the \citet{tappin2022} HI1A broadband photometric calibration is indicated by the dotted line and shaded region. The uncertainties in these values were generously estimated by the scatter of the HI1A points, allowing for potentially correlated measurement errors comparable to this scatter. Note that \ion{Na}{1} D sensitivity varies with the assumed $r=1$~au ionization rate, with the values in this plot computed for an assumed rate of $7.59\times10^{-6}$~s$^{-1}$ \citep{huebner2015}.}
\label{fig:merc}
\end{figure*}

\begin{figure}
\centering
\includegraphics[width=\linewidth]{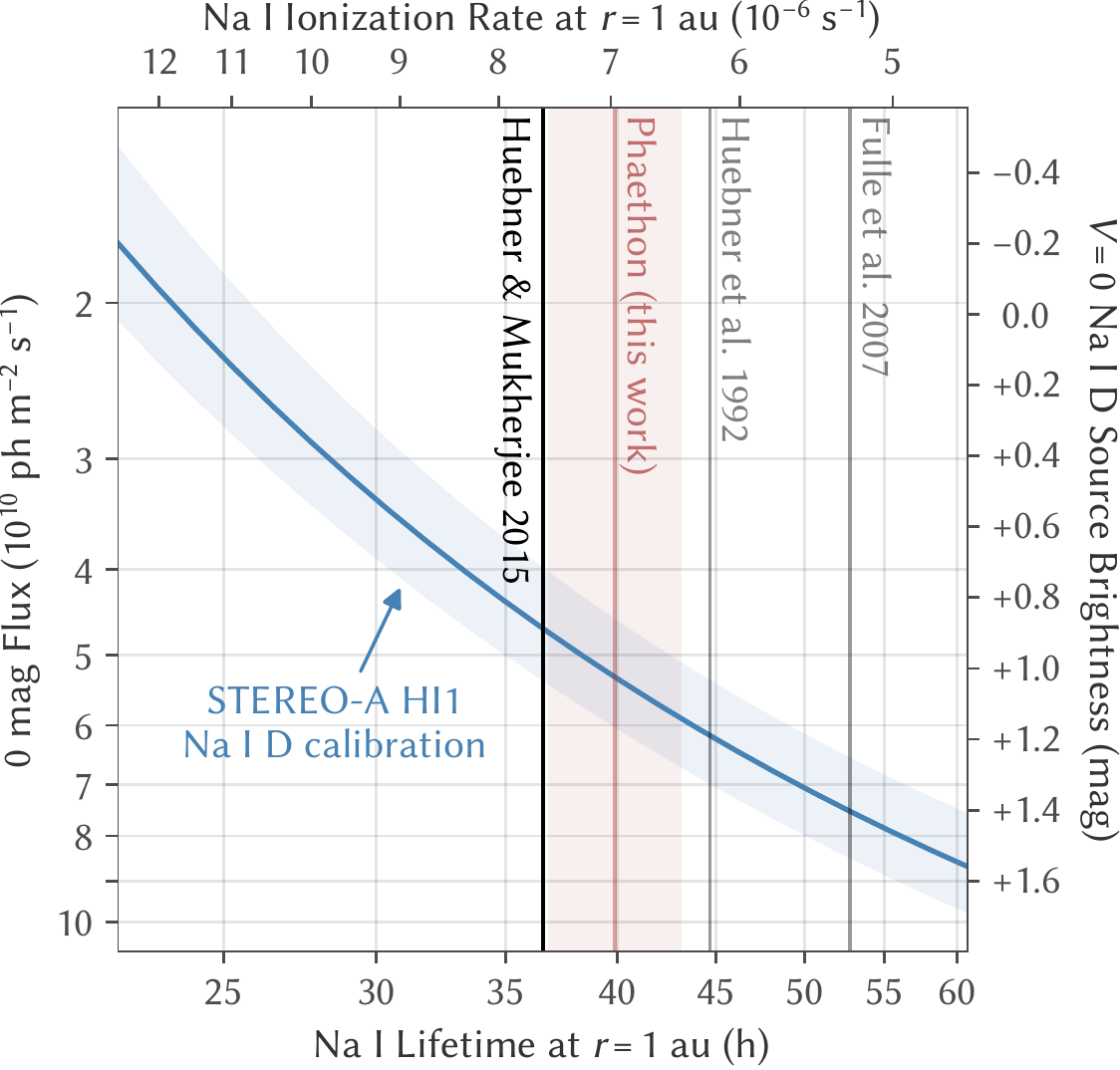}
\caption{Variation of the STEREO-A HI1 \ion{Na}{1} D calibration with assumed \ion{Na}{1} lifetime at $r=1$~au (with $\propto r^2$ scaling). Values from \citet[quiet and active Sun mean; selected as our nominal a priori value]{huebner2015}, \citet[quiet and active Sun mean]{huebner1992}, and \citet{fulle2007} are indicated by their labeled vertical lines. We also indicate the $r=1$~au equivalent $(40\pm3)$~h lifetime from our Phaethon photometric fit, for comparison.}
\label{fig:merc_lt}
\end{figure}

Like Phaethon, Mercury has also been known to generate a tail of \ion{Na}{1} accelerated by radiation pressure, albeit with atoms likely sourced from predominantly non-thermal processes \citep{potter2002}. Following reports of this tail's unexpectedly high brightness in HI1 imagery under the preflight transmission profiles \citep{schmidt2010b}, \citet{halain2012} retested the HI1 engineering qualification model in 2010 and found that the filter bandpass had not only degraded from aging since the initial measurements in 2005, but was also affected by vacuum and the low operational temperatures. These effects combined to produce a net $\sim$20~nm blueward shift in the filter transmission that still left the 589.0/589.6~nm \ion{Na}{1} D lines just beyond the new $\sim$595--720~nm FWHM interval, but substantially raised their relative transmission by an order of magnitude from 1--2\% to $\sim$15\% that is sufficient to attribute the observed brightness of Mercury's tail to \ion{Na}{1} D emission alone. The next brightest species seen in Mercury's tail at optical wavelengths is \ion{K}{1} \citep{lierle2022}, but its equivalent 766.5/769.9~nm resonance lines have $<$1\% relative transmission through HI1, while \ion{K}{1} is produced at only $\sim$1\% the rate of \ion{Na}{1} and photoionizes more rapidly \citep{huebner2015}, so negligibly contributes to the tail's brightness in HI1 imagery.

Both of the STEREO HI instruments have high precision, time-dependent photometric calibrations available for broadband source photometry derived by monitoring a sample of stars over the course of the mission \citep{tappin2017,tappin2022}. These calibrations, however, are only minimally sensitive to the exact transmission profiles. As the \ion{Na}{1} D lines fall near the edge of the HI1 bandpass, HI1's sensitivity to them is strongly dependent on the precise bandpass shift and could theoretically vary substantially over the mission lifetime. We therefore opted to perform a separate calibration of HI1's sensitivity to pure \ion{Na}{1} D emission using Mercury's tail as a flux standard.

\citet{schmidt2010a} measured Mercury's \ion{Na}{1} escape rate to be $Q(\text{\ion{Na}{1}})=1.26\times10^{24}$~atoms~s$^{-1}$ at a true anomaly of $\theta=68^\circ$ near the seasonal peak in radiation pressure and $Q(\text{\ion{Na}{1}})\approx5\times10^{23}$~atoms~s$^{-1}$ at $\theta\approx115^\circ$ using a similar tail model as the one we describe in Appendix~\ref{sec:natail} and used for Phaethon. We assume the tail brightness at these $\theta$ remains similar at every orbit, and select the HI1 data near these $\theta$ for flux calibration. However, light scattered by Mercury's bright, daylit surface introduces diffraction and saturation artifacts in HI1 imagery that complicates measurements of the comparatively faint \ion{Na}{1} tail. We limited our analysis to epochs where Mercury is observed at high phase angles $\alpha>145^\circ$ to minimize the light from the daylit surface and simultaneously maximize the observed \ion{Na}{1} column density, as the tail is projected nearly along the line of sight in this geometry.

We selected seven epochs from 2007 to 2021 that met our criteria for analysis, as listed in Table~\ref{tab:mercobs}. At each epoch, we stacked all frames from a 2--3~day period, and show a cutout of the tail at every epoch in Figure~\ref{fig:merc}. We then extracted the linear brightness profile of the tail out to $1^\circ$ using $8'$ wide rectangular photometric apertures extending from Mercury in the antisunward direction, and using the $8'$ on both sides of this tail region for background determination. We excluded the innermost $10'$ of the tail, where artifacts associated with light from the surface are noticeable. Next, we used a quadratic polynomial fit of \citet{schmidt2010a}'s \ion{Na}{1} escape rates with respect to $\theta$ to determine the expected $Q(\text{\ion{Na}{1}})$ at each epoch, which we estimated as being $\pm10\%$ from the true value. We then computed the expected \ion{Na}{1} D tail profiles with our \ion{Na}{1} tail model, and fitted them with the observed tail profiles to determine the \ion{Na}{1} D sensitivity of the instruments.

We provide the computed sensitivities for a nominal $r=1$~au \ion{Na}{1} photoionization rate of $7.59\times10^{-6}$~s$^{-1}$ \citep{huebner2015} in Table~\ref{tab:mercobs} and plot them in Figure~\ref{fig:merc}. Our measurements appear consistent with a fixed 0~mag flux of $(4.7\pm0.7)\times10^{10}$~ph~m$^{-2}$~s$^{-1}$ over the full 2007--2021 calibration period for STEREO-A HI1. The single measurement for STEREO-B HI1 in 2010 suggests this camera has comparable \ion{Na}{1} sensitivity. This result indicates the shift in bandpass had likely already completed by the commencement of STEREO mission operations, so we consider the relative \ion{Na}{1} D sensitivity to be constant over the operating lifetimes of both cameras.

Additionally, all of our tail brightness measurements are weighted toward distant portions of Mercury's tail where the fraction of surviving \ion{Na}{1} and thus the relative brightness of the tail are sensitive to the assumed \ion{Na}{1} lifetime. Earlier estimates by \citet{huebner1992} and \citet{fulle2007} differ from our nominal value by tens of percent, which would shift the calibrated \ion{Na}{1} D sensitivity beyond our stated uncertainties. Figure~\ref{fig:merc_lt} shows the variation in the STEREO-A HI1 \ion{Na}{1} D calibration across a range of assumed \ion{Na}{1} lifetime, which we incorporated as an \ion{Na}{1} lifetime-dependent calibration into our photometric model to address this concern.

The final fit to Phaethon's photometry in Section~\ref{sec:nafit} provided an equivalent $r=1$~au \ion{Na}{1} lifetime of $(40\pm3)$~h, corresponding to a photoionization rate of $(7.0\pm0.6)\times10^{-6}$~s$^{-1}$ that is close to the selected \citet{huebner2015} value. This retrospective analysis scales the 0~mag flux to $(5.3\pm0.7)\times10^{10}$~ph~m$^{-2}$~s$^{-1}$, which remains consistent with the a priori value.

\section{LASCO Photometric Calibration}
\label{sec:lascophot}

We consider LASCO observations spanning its multi-decade lifetime, over which variations in detector sensitivity could significantly affect photometry. Existing calibrations of both LASCO C2 \citep{llebaria2006,gardes2013,colaninno2015} and C3 \citep{thernisien2006} cover only a small fraction of the mission lifetime and/or do not calibrate filters used by our observations. We therefore opted to perform a photometric calibration encompassing the entire time period to ensure the validity of our photometry.

By convention, SOHO and STEREO zero-point magnitudes are set to provide the Johnson $V$ magnitudes of solar-colored stars, which defines a magnitude system where the mean magnitude of the Sun from 1~au is $-26.76$ in all bandpasses \citep{willmer2018}. We based our calibration on a single, high quality flux standard star to minimize complications with variability and color transformations. We chose to use 39~Tau due to its brightness ($V=5.90$), its minimal ($<$0.1~mag) variability \citep{gaia2022}, its nearly solar optical colors \citep{farnham2000}, and its proximity to the ecliptic plane that causes it to transit both the C2 and C3 fields of view over several days each May.

We processed all frames containing 39~Tau over 1996--2022 in the same manner as the science data, producing a yearly median stack for every filter with at least three frames present, and performing aperture photometry within a $45''$ radius for C2 and a $2'$ radius for C3. Under the synoptic programs in which these data were collected, only C2 Orange and C3 Clear frames are collected at full resolution, with all other frames recorded with $2\times2$ binning yielding a slightly larger effective PSF. To mitigate this effect, we corrected the measured fluxes to larger apertures of $2'$ in C2 and $5'$ in C3, which we treat as equivalent to an infinite aperture fully capturing the flux of a point source. We measured the C2 PSF from 39~Tau itself and found a minimal correction factor of $1.01\times$ regardless of binning. However, we measured the C3 PSF from the neighboring star 37~Tau due to the presence of a $V=8$ star ${\sim}3'$ northeast of 39~Tau that prevents the use of apertures much larger than our $2'$ radius, and found correction factors of $1.1\times$ for unbinned frames and $1.3\times$ for $2\times2$ binned frames.

\begin{figure}
\centering
\includegraphics[width=\linewidth]{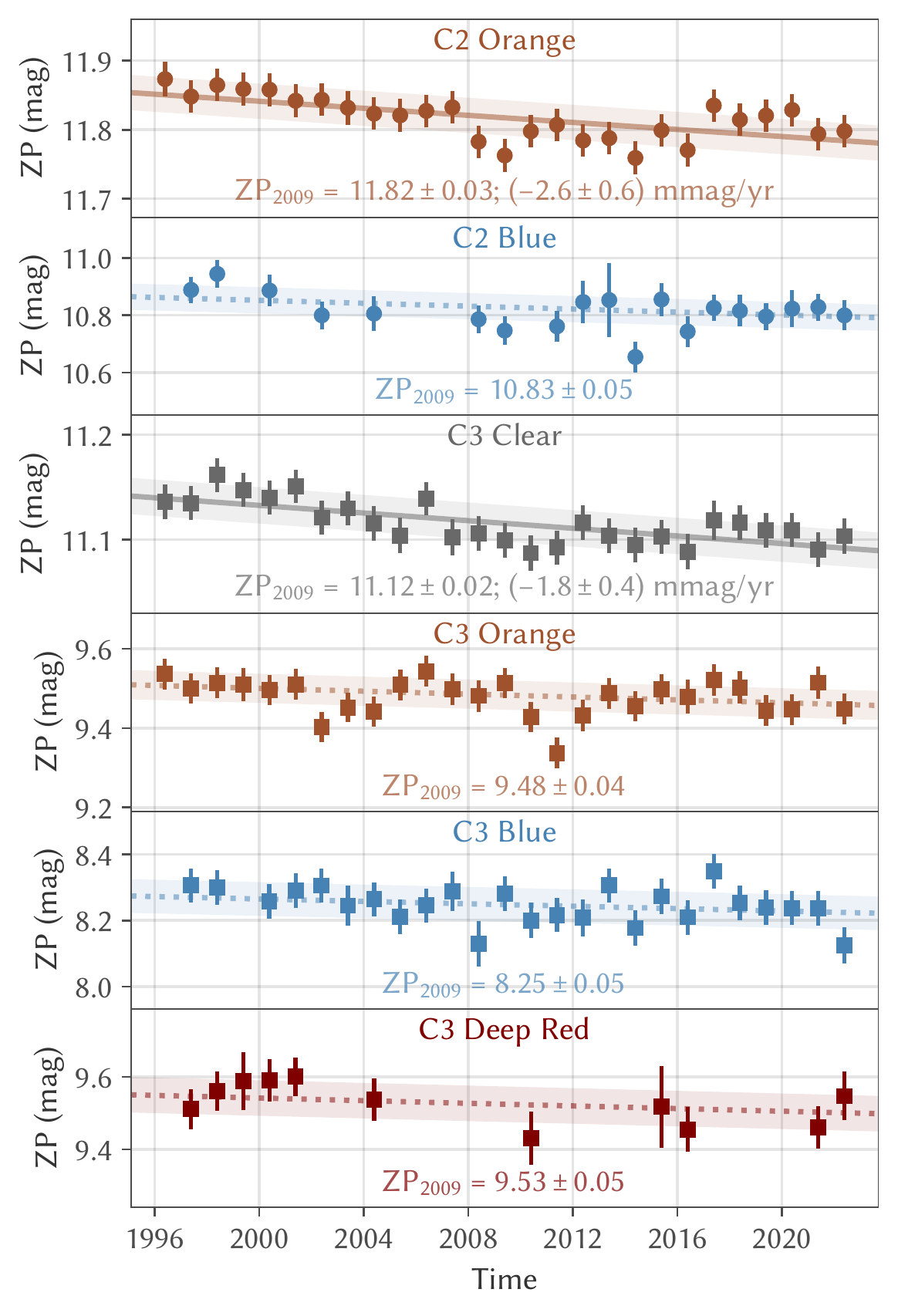}
\caption{LASCO C2 and C3 zero-point (ZP) magnitudes derived from 39~Tau. The lines and associated shading indicate linear fits to the zero-points, with the slopes in all C2 and C3 assumed to match the fitted slopes of C2 Orange and C3 Clear, respectively.}
\label{fig:39tau}
\end{figure}

Figure~\ref{fig:39tau} plots these resulting zero-points which are generally consistent with the previously published calibrations over their respective time periods, along with a gradual sensitivity decline of ($-2.6\pm0.6$)~mmag~yr$^{-1}$ in C2 Orange and ($-1.8\pm0.4$)~mmag~yr$^{-1}$ in C3 Clear, which we adopt as the rates for all C2 and C3 bandpasses, respectively. For comparison, \citet{llebaria2006} measured $-8$~mmag~yr$^{-1}$ for C2 Orange over 1996--2004, and \citet{thernisien2006} measured $(-4.8\pm1.1)$~mmag~yr$^{-1}$ for C3 Clear over 1996--2003. A subsequent recalibration of C2 Orange by \citet{gardes2013} measured $-3.8$~mmag~yr$^{-1}$ over 1999--2009, while \citet{colaninno2015} measured $(-2.2\pm0.3)$~mmag~yr$^{-1}$ over 1996--2013, the latter comparable to our 1996--2022 result.

\newcolumntype{C}{@{\extracolsep{2mm}}c@{\extracolsep{2mm}}}
\begin{deluxetable*}{llCcCcCc}
\tablecaption{SOHO LASCO/STEREO HI1 Bandpass Photometric Properties}
\label{tab:lascophot}

\tablecolumns{8}
\tablehead{
\multicolumn{4}{c}{} & \multicolumn{4}{c}{\ion{Na}{1} D Calibration} \\
\cline{5-8}
\multicolumn{2}{c}{Bandpass} & \multicolumn{2}{c}{Photometric Zero-point\tablenotemark{a}} & \multicolumn{2}{c}{A Priori\tablenotemark{b}} & \multicolumn{2}{c}{A Posterori\tablenotemark{c}} \\
\cline{1-2} \cline{3-4} \cline{5-6} \cline{7-8}
\colhead{Instrument} & \colhead{Filter} & \colhead{At 2009.0} & \colhead{Slope} & \colhead{0~mag Flux} & \colhead{$V=0$} & \colhead{0~mag Flux} & \colhead{$V=0$} \\
\colhead{} & \colhead{} & \colhead{(mag)} & \colhead{(mmag yr$^{-1}$)} & \colhead{(ph~m$^{-2}$~s$^{-1}$)\tablenotemark{d}} & \colhead{(mag)\tablenotemark{e}} & \colhead{(ph~m$^{-2}$~s$^{-1}$)\tablenotemark{d}} & \colhead{(mag)\tablenotemark{e}}
}

\startdata
SOHO LASCO C2 & Orange & $11.81\pm0.02$ & $-2.8\pm0.6$ & $8.6\times10^9$ & $-0.94$ & $(8.7\pm1.3)\times10^9$ & $-0.94\pm0.16$ \\
& Blue & $10.81\pm0.06$ & & \nodata & \nodata & \nodata & \nodata \\
SOHO LASCO C3 & Clear\tablenotemark{f} & $11.11\pm0.02$ & $-1.9\pm0.4$ & $(2.9\pm0.2)\times10^{10}$ & $+0.39\pm0.05$ & $(2.7\pm0.4)\times10^{10}$ & $+0.31\pm0.16$ \\
& Orange & $9.48\pm0.04$ & & $8.6\times10^{9\phantom{0}}$ & $-0.94$ & $(8.7\pm1.3)\times10^9$ & $-0.94\pm0.16$ \\
& Blue & $8.24\pm0.05$ & & \nodata & \nodata & \nodata & \nodata \\
& Deep Red\tablenotemark{g} & $9.53\pm0.05$ & & \nodata & \nodata & \nodata & \nodata \\
STEREO HI1\tablenotemark{h} & \nodata & \nodata & \nodata & $(4.7\pm0.7)\times10^{10}$ & $+0.89\pm0.16$ & $(5.3\pm0.7)\times10^{10}$ & $+1.02\pm0.15$ 
\enddata

\tablenotetext{a}{Fitted zero-point linear model, with intercept at year 2009.0. Slopes fitted only for C2 Orange and C3 Clear, and assumed to be identical for the other C2 and C3 bandpasses.}
\tablenotetext{b}{Relative sensitivity to \ion{Na}{1} D emission expected from LASCO bandpass profiles and HI1 Mercury tail observations.}
\tablenotetext{c}{\ion{Na}{1} D sensitivity metrics corrected from fit to Phaethon's photometry.}
\tablenotetext{d}{\ion{Na}{1} D flux producing magnitude 0.}
\tablenotetext{e}{Magnitude of a Johnson $V=0$ source emitting only \ion{Na}{1} D.}
\tablenotetext{f}{A priori \ion{Na}{1} D calibration based on assumed bandpass, as described in text.}
\tablenotetext{g}{Not used for Phaethon observations; included for reference only.}
\tablenotetext{h}{From STEREO-A HI1 calibration in Appendix~\ref{sec:hi1sens}; STEREO-B treated as identical.}
\end{deluxetable*}
\vspace{-2.5em}

Unlike HI1, LASCO cannot readily detect Mercury's \ion{Na}{1} tail due to its lower instrumental sensitivities and the brighter coronal background within its fields of view. However, none of the LASCO filters have a bandpass cut-on or -off near the \ion{Na}{1} D lines, so minor shifts in bandpass like those of HI1 will not significantly alter the \ion{Na}{1} D sensitivity. The preflight filter transmission profiles show that only C2 Orange, the nearly identical C3 Orange, and C3 Clear transmit \ion{Na}{1} D emission, with 589~nm near the peak transmission of all three filters.

However, the cut-on wavelength of C3 Clear is below 500~nm where no transmission or detector quantum efficiency (QE) data appear to be available for C3. Given the similarities of the C2 and C3 detectors and their measured $>$500~nm QE profiles, we use the C2 detector QE as that of the C3 detector below 500~nm, and use the mean 500--510~nm transmission of C3 Clear filter as its assumed transmission at $<$500~nm to obtain an initial estimate for the full C3 Clear bandpass.

This estimated C3 Clear bandpass is $5.5\times$ more sensitive to a solar-colored source than C3 Orange, corresponding to a 1.86~mag higher zero-point magnitude for C3 Clear. However, the measured zero-point of C3 Clear is only $(1.64\pm0.05)$~mag higher than that of C3 Orange, or ${\sim}4.5\times$ greater solar sensitivity in the former. Even 0\% transmission below 500~nm yields too high of a solar sensitivity, suggesting the transmission longward and/or shortward of the Orange bandpass interval is lower than predicted from preflight transmission and QE alone, possibly from the LASCO optics. The precise profile, however, is not required for quantifying \ion{Na}{1} D sensitivity. We assume the central portion of the Clear profile---including the Orange bandpass interval---remains fixed, then compensate for the $(0.22\pm0.05)$~mag discrepancy in zero-point color by contracting the effective Clear bandpass by $(18\pm4)\%$ (i.e., lowering the Clear sensitivity anywhere outside the Orange bandpass interval to reach the observed zero-point color).

\ion{Na}{1} D emission with a Johnson $V$ magnitude of 0 would appear as magnitude $+0.39\pm0.05$ in this corrected C3 Clear bandpass, or $\sim$1.3~mag fainter than the $-0.94$ expected for the C2 and C3 Orange bandpasses. Table~\ref{tab:lascophot} provides the final solar and \ion{Na}{1} D photometric calibration parameters for LASCO C2/C3, as well as the HI1 \ion{Na}{1} D sensitivity obtained in Appendix~\ref{sec:hi1sens} for comparison. We also include a posteriori values for \ion{Na}{1} D calibration parameters, which have been corrected for the \ion{Na}{1} lifetime and color offsets fitted from Phaethon's photometry in Section~\ref{sec:nafit}, although all values remain consistent with the a priori calibration.

\section{Thermal Desorption of Sodium}
\label{sec:desorp}

In order to extrapolate Phaethon's \ion{Na}{1} production to $r<0.14$~au for comparison with 322P, 323P and other sunskirting asteroids, we consider a rudimentary sublimation model for a fast rotating, isothermal blackbody asteroid following the approaches of \citet{huebner1970}, \citet{sekanina2003}, and \citet{cranmer2016}. From the Clausius--Clapyeron relation, the vapor pressure over a surface of temperature $T$ comprised of a substance with a unit latent heat of sublimation $L$ is

\begin{equation}
P_v(T)=P_0\exp\left(-\frac{L}{k_B}\left(\frac{1}{T}+\frac{1}{T_0}\right)\right)
\end{equation}

\noindent where $k_B$ is the Boltzmann constant, $P_0\equiv398$~MPa, and $T_0$ is a material-dependent temperature with $T_0\sim5000$~K for typical refractory, asteroidal materials \citep{cranmer2016}. For metallic Na, $L=1.097$~eV~atom$^{-1}=105.8$~kJ~mol$^{-1}$ and $T_0=4660$~K \citep{huebner1970}.

Phaethon's surface, however, is likely not comprised of a single substance, but instead of a number of different materials spanning a wide range of volatility each with their own unique vapor pressure functions. For simplicity, we assume the observed \ion{Na}{1} originates exclusively from a single volatile substance---which we assume is pure Na, but could in theory be an Na-bearing compound that only later produces \ion{Na}{1} upon dissociation---that is embedded within an otherwise perfectly refractory substrate that contributes no additional vapor pressure.

Suppose momentarily that this Na were segregated in large, macroscopic clumps covering a fraction $f_\mathrm{eff}$ of the surface, and thus occupying a volume fraction $f_\mathrm{eff}$ of the surface material. The effective vapor pressure over the surface---and thus the expected \ion{Na}{1} production rate---would be $f_\mathrm{eff}$ times that of if the volatile covered the full surface. In reality, the Na is unlikely be segregated in this manner, and is most likely more finely divided throughout the refractory substrate to which the atoms are largely individually bonded. Moreover, the steep dependence of vapor pressure and thus Na loss on temperature likely causes the surface material within the upper few centimeters, which reaches subsolar temperatures of $\sim$1000~K near perihelion, to fully devolatilize within hours of exposure to such temperatures, as \citet{masiero2021} experimentally demonstrated for chrondritic meteorites serving as analogs to Phaethon's surface.

As we have found no evidence for ongoing resurfacing activity clearing the devolatilized surface material, we presume the sources of Na responsible for the observed activity are sequestered some distance beneath the surface where cooler peak temperatures have sufficiently slowed their depletion for their continued presence. We model the subsurface Na abundance as a step function, with 0\% Na in the devolatilized layer and a constant effective Na fraction $f_0$ beneath. We found in Section~\ref{sec:nafit} that the peak in Phaethon's \ion{Na}{1} production lagged its perihelion by $(3.0\pm0.8)$~h, a value comparable to its 3.6~h rotation period; if fully attributed to the thermal propagation time into the active subsurface layer, the observed offset suggests the devolatilized layer extends to roughly the diurnal skin depth, a thickness of $\Delta z\sim0.1$~m.

Under this revised picture, $f_\mathrm{eff}$ evidently no longer measures the now $\sim$0\% surface Na content, but still serves as a proxy for subsurface volatility that encapsulates both the actual subsurface Na abundance below the devolatilized layer and the resistance of this devolatilized layer to the diffusion of \ion{Na}{1} through it. We use the model by \citet{schorghofer2008} for the sublimation through a medium of porosity $\phi\sim0.5$, grain radii $R_g\sim0.1$~mm, and tortuosity $\tau_\mathrm{tort}\sim1$, as assumed by \citet{masiero2021}, and find the effect of the devolatilized layer on Na desorption below to be

\begin{equation}
\frac{f_\mathrm{eff}}{f_0}\approx\frac{4\pi\phi R_g}{(8+\pi)(1-\phi)\tau_\mathrm{tort}\Delta z}\sim10^{-3}
\label{eq:barrier}
\end{equation}

\noindent indicating that such a devolatilized layer can indeed significantly suppress Na loss from below. We caution, however, that the lack of reliable constraints on the input parameters limits this estimate to an order of magnitude uncertainty at best.

\edit{1}{One additional consequence of the Na being largely sequestered beyond the diurnal skin depth is that Na desorption should have little diurnal variation, as the temperature of the Na-bearing subsurface remains near the diurnally-averaged surface temperature at all times. As Phaethon is near equinox at perihelion, its subsurface temperature there and thus \ion{Na}{1} production likely exhibits only minimal day--night asymmetry, although the latitudinal variation in diurnally-averaged insolation will still restrict the strongly temperature-dependent Na desorption to a fraction $f_T<1$ of the surface at any instant.}

\edit{1}{At its fitted $n=13.7\pm0.5$, Phaethon's \ion{Na}{1} production falls off to 1/2 the peak rate at $1.05\times$ its perihelion distance where solar flux is $1.05^{-2}=0.90\times$ the perihelion value. If this variation with incident solar flux holds locally across the entire surface, then local \ion{Na}{1} loss near perihelion/equinox should fall to 1/2 the equatorial rate at roughly $\pm25^\circ$ latitude where insolation is $\cos(25^\circ)\approx0.90\times$ the equatorial value, bounding an equatorial region with $>$1/2 the peak \ion{Na}{1} flux that occupies a fraction $f_T\approx0.42$ of Phaethon's surface. This fraction depends slightly on $n$---increasing to just $f_T\approx0.52$ at 322P's lower $n=8.8\pm0.3$---as well as on less well-constrained properties including local variation in subsurface Na content/depth, deviations from the modeled spherical shape, and associated seasonal effects.}

\edit{1}{We then treat this equatorial region of area $4\pi R^2 f_T$ as the only active portion of the radius $R$ asteroid, which we model as having a uniform subsurface temperature $T_\mathrm{Na}$ equal to the temperature of the desorbed \ion{Na}{1}.} The Hertz--Knudsen equation then gives

\begin{equation}
\frac{Q(\text{\ion{Na}{1}})}{4\pi R^2}=\frac{\alpha_s f_T f_\mathrm{eff} P_v(T_\mathrm{Na})}{\sqrt{2\pi m_\mathrm{Na}k_B T_\mathrm{Na}}}
\label{eq:prodrate}
\end{equation}

\noindent where $m_\mathrm{Na}=23$~u is the atomic mass of Na, and $\alpha_s$ is the sublimation efficiency which we treat as roughly unity, as typical of atoms and small molecules \citep{vanlieshout2014}. \edit{1}{Thermal radiation from the asteroid is set by the actual surface temperature $T_s$ through the Stefan--Boltzmann law integrated over its surface $A$.} The energy balance between the incident solar flux $S(r)=(1361~\mathrm{W~m^{-2}})\times(1~\mathrm{au}/r)^2$ \citep{prvsa2016} with the outgoing thermal radiation and the sublimation power then gives

\begin{equation}
\begin{aligned}
(1-A_B)S(r)\times\pi R^2&=\varepsilon\sigma_\mathrm{SB}\oiint_A T_s^4 dA'+L\times Q(\text{\ion{Na}{1}})\\
&\equiv\varepsilon\sigma_\mathrm{SB}T_\mathrm{eff}^4\times4\pi R^2+L\times Q(\text{\ion{Na}{1}})
\end{aligned}
\end{equation}

\noindent for which we choose a convenient Bond albedo $A_B=0.1$ and effective emissivity $\varepsilon=0.9$ while neglecting surface roughness and thermal inertia, \edit{1}{and defining an effective temperature $T_\mathrm{eff}$ as the temperature of an isothermal sphere equal in size to and radiating the same power as the actually non-isothermal asteroid}. In practice, we found that the $Q(\text{\ion{Na}{1}})$ of Phaethon, and even of 322P and 323P, are so small that the sublimation term $L\times Q(\text{\ion{Na}{1}})$ actually contributes negligible cooling for all plausible $L$, so $T_\mathrm{eff}$ becomes well-approximated at all relevant $r$ by the standard isothermal blackbody temperature

\begin{equation}
\label{eq:isobb}
T_\mathrm{eff}(r)\approx\left(\frac{S(r)}{4\sigma_\mathrm{SB}}\right)^{1/4}\equiv T_1\times\sqrt{\frac{1~\mathrm{au}}{r}}
\end{equation}

\noindent with $T_1=278$~K, giving $T_\mathrm{eff}(0.14~\mathrm{au})=743$~K at Phaethon's perihelion. \edit{1}{We then approximate the subsurface temperature $T_\mathrm{Na}$ as the surface temperature $T_s$ averaged along the equator (i.e., the diurnally-averaged temperature at any point on the equator). The relation between $T_\mathrm{Na}$ and $T_\mathrm{eff}$ is therefore dependent on the thermal properties of the surface material. For a crude, NEATM-like $T_s\propto(\cos\psi)^{1/4}$ at solar zenith angles $\psi<90^\circ$ that neglects nightside surface temperature \citep{harris1998}, we find $T_\mathrm{Na}\approx0.6T_\mathrm{eff}$. However, thermophysical modeling of Phaethon by both \citet{yu2019} and \citet{masiero2021} found equatorial nightside surface temperatures near perihelion to be a non-negligible $\sim$50\% of the subsolar temperature ($\sim$1000~K), which raises the diurnally-averaged equatorial temperature to a much higher $T_\mathrm{Na}\approx T_\mathrm{eff}$. We therefore adopt this value as our nominal estimate for the temperature driving Na desorption.}

\edit{1}{With their model, \citet{yu2019} actually computed slightly higher peak diurnally-averaged surface temperatures of $\sim$800~K. \citet{masiero2021}, meanwhile, directly computed the perihelion temperature of material $\sim$0.1~m beneath the surface at near-equatorial latitudes and found it to be a slightly cooler $\sim$600--700~K, possibly due in part to thermal lag shifting the peak subsurface temperatures to a few hours after perihelion. Given these comparisons, we consider our $T_\mathrm{Na}\approx T_\mathrm{eff}$ approximation likely accurate to $\pm$100~K at Phaethon's perihelion, corresponding to $\pm$37~K in $T_1$.}

With this subsurface temperature model, \ion{Na}{1} production scales with $r$ as

\begin{equation}
Q(\text{\ion{Na}{1}})\propto r^{1/4}\exp\left(-\frac{L}{k_B T_1}\times\sqrt{\frac{r}{1~\mathrm{au}}}\right)
\end{equation}

The logarithmic slope of \ion{Na}{1} with $r$ then becomes a pure function of $r$ and $L$:

\begin{equation}
n\equiv-\frac{\partial\ln Q}{\partial\ln r}=\frac{L}{2k_B T_1}\times\sqrt{\frac{r}{1~\mathrm{au}}}-\frac{1}{4}
\end{equation}

Phaethon's fitted $n=13.7\pm0.5$ is considerably steeper than the $n=9.6\pm0.8$ expected for metallic Na at $r=0.14$~au, and instead corresponds to Na more tightly bound with $L=(1.8\pm0.2)~\mathrm{eV~atom^{-1}}=(170\pm20)~\mathrm{kJ~mol}^{-1}$. \edit{1}{Phaethon's $Q(\text{\ion{Na}{1}})=(5.5\pm0.8)\times10^{23}$~atoms~s$^{-1}$ at $r=0.14$~au then corresponds to an effective Na fraction of $f_\mathrm{eff}=10^{-4.5\pm0.5}$ by Equation~\eqref{eq:prodrate}---equivalent to a subsurface Na fraction $f_0\sim10^{-2}$--$10^{-1}$ beneath the $\Delta z\sim0.1$~m thick devolatilized layer by Equation~\eqref{eq:barrier}, which we consider broadly consistent with the $\sim$0.5\% Na mass abundance typical of chondritic meteorites \citep{lodders2021} given the abundance of poorly-constrained parameters contained in these estimates.}

This model can also set a rough bound on the age of Phaethon's surface $t_1$ since it was last resurfaced (i.e., when $f_\mathrm{eff}/f_0\approx1$). Treating $f_0$ as approximately an Na mass fraction, Phaethon's current orbitally averaged \ion{Na}{1} loss rate $\bar{Q}=(1.05\pm0.15)\times10^{29}$~atoms~orbit$^{-1}=(7.3\pm1.0)\times10^{28}$~atoms~yr$^{-1}$ deepens the devolatilized layer at a rate

\begin{equation}
\Delta\dot{z}\sim\frac{\bar{Q}m_\mathrm{Na}}{4\pi R^2\rho f_0}\sim10^{-14}~\mathrm{m}~\mathrm{s}^{-1}
\label{eq:deplete}
\end{equation}

\noindent for Phaethon's radius $R=2.7$~km and bulk density $\rho=1.6$~g~cm$^{-3}$ \citep{maclennan2021}.

More generally, $\bar{Q}\propto f_\mathrm{eff}$ at constant $T$, by Equation~\eqref{eq:prodrate}, and $f_\mathrm{eff}\propto1/\Delta z$, by Equation~\eqref{eq:barrier}, and $\Delta\dot{z}\propto\bar{Q}\propto1/\Delta z$. Then, after integrating time $0\to t_1$ corresponding to $f_\mathrm{eff}$ from $f_0$ to its current $\sim10^{-3}f_0\ll f_0$:

\begin{equation}
\Delta z\propto t_1^{1/2}\implies\frac{\Delta\dot{z}}{\Delta z}\sim\frac{1}{2t_1}
\end{equation}

Then, $t_1\sim\Delta\dot{z}/(2\Delta z)\sim10^4$~yr for the current values of $\Delta\dot{z}$ and $\Delta z$. In reality, the Na increasingly deviates from the isothermal blackbody temperature at lower $\Delta z$, where Na is closer to the surface and thus experiences more extreme diurnal temperature fluctuations. Under those past conditions, $\bar{Q}$ and thus $\Delta\dot{z}$ would both be higher than these scaling approximation imply, so the computed $t_1$ actually reflects a rough upper bound on the surface age of ${\lesssim}10^4$~yr.

We caution that this age may not necessarily be the actual time since the surface was last cleared, but rather the cumulative duration over which the surface has experienced a comparable degree of heating as on its present orbit. Phaethon has only spent $\sim$10~kyr total at $q\lesssim0.15$~au over the past 100~kyr, so our ${\lesssim}10^4$~yr bound remains insufficient to distinguish a true resurfacing event, as might be associated with the Geminids formation, from the brevity of Phaethon's sunskirting history. More useful bounds require higher precision calculations, likely from proper thermophysical and dynamical modeling in conjunction with better constrained surface material properties.

\section{Sodium Tail Model}
\label{sec:natail}

\begin{figure*}
\centering
\includegraphics[width=\linewidth]{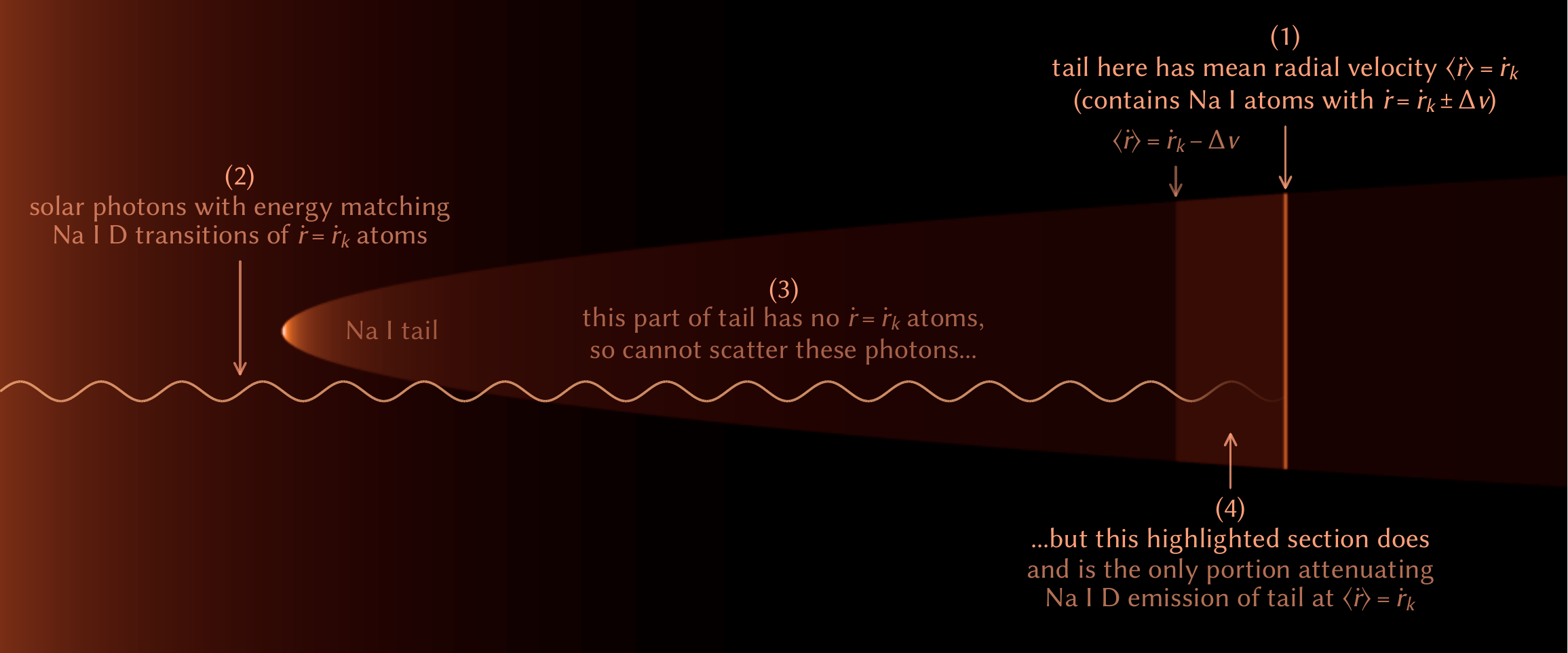}
\caption{Illustration explaining the \ion{Na}{1} tail model's simplified treatment of optical depth along the tail, where only the \ion{Na}{1} atoms immediately sunward of any point along the tail contribute to the attenuation of \ion{Na}{1} D emission at that point.}
\label{fig:depth}
\end{figure*}

\restartappendixnumbering

Here, we describe a numerical model for the brightness profile of a predominantly optically thin \ion{Na}{1} tail, which we used to quantify the \ion{Na}{1} D emission of Phaethon and other sunskirting objects.

We begin with a fully optically thin tail model, similar to those previously used by \citet{cremonese1997} and \citet{schmidt2010a}. We model \ion{Na}{1} fluorescence with two independent two-level systems corresponding to the 589.0~nm D$_2$ and 589.6~nm D$_1$ transitions, and treat the instantaneous fraction of excited atoms as zero, neglecting the impact of other transitions, as appropriate for \ion{Na}{1} excited only by stimulated absorption of solar photons. We use \ion{Na}{1} D transition parameters from \citet{sansonetti2008} and the disk-averaged solar spectrum from \citet{neckel1999} to compute the corresponding fluorescence efficiencies (emission rates per atom) as functions of heliocentric distance $r$ and radial velocity $\dot{r}$, capturing the Swings effect \citep{swings1941}. We correct for anisotropic scattering by the D$_2$ line by multiplying its contribution to the flux observed at phase angle $\alpha$ by $0.967\times(1+0.102\cos^2\alpha)$ \citep{chamberlain1961}.

Next, we numerically integrate the motion of an atom released at initial $r=r_0$ and $\dot{r}=\dot{r}_0$, and accelerated antisunward at $\ddot{r}$ by the momentum transfer of $\mathrm{D}_2+\mathrm{D}_1$ absorption. The linear tail brightness profile for an \ion{Na}{1} source at this $r_0$ and $\dot{r_0}$ then corresponds to the emission rate at the $r$ and $\dot{r}$ along the tail, divided by $\dot{r}-\dot{r}_0$ into an emission rate per unit length, and modulated by the photoionization of \ion{Na}{1}. We initially used a photoionization rate of  $(7.59\times10^{-6}~\mathrm{s}^{-1})\times(1~\mathrm{au}/r)^2$ from the mean of the quiet and active Sun rates found by \citet{huebner2015}, but also later directly fitted for (the reciprocal of) this value in Section~\ref{sec:nafit}.

We crudely estimate the \ion{Na}{1} outflow speed $v_\mathrm{out}$ as the thermal speed $(8k_B T(r)/(\pi m_\mathrm{Na}))^{1/2}$ for the isothermal blackbody temperature $T(r)$ from Equation~\eqref{eq:isobb}. These values theoretically influence the morphology of an optically thin \ion{Na}{1} tail in two distinct ways, although neither actually substantially affects our modeling of LASCO/HI1 observations of sunskirting objects:

\begin{enumerate}
\item A population of \ion{Na}{1} with a broadly distributed $\dot{r}$ around a mean $\langle\dot{r}\rangle$ will sample a broad region of the solar spectrum, and thus fluoresce with different efficiency than a different population of \ion{Na}{1} where all atoms have the $\dot{r}=\langle\dot{r}\rangle$. We approximated this effect by taking an ensemble average of tail profiles with initial $\dot{r}$ normally distributed about the nominal $\dot{r}_0$ with standard deviation $v_\mathrm{out}$, but found its impact to be minimal at the $v_\mathrm{out}\sim$1~km~s$^{-1}$ of sunskirting objects. The solar spectrum near the \ion{Na}{1} D lines is fairly smooth at $\sim$1~km~s$^{-1}$ scales, and requires much a wider spread in $\dot{r}$ before any difference in fluorescence efficiency becomes noticeable.
\item The outflow speed becomes the expansion rate of \ion{Na}{1} parcels propagating down the tail, which sets the tail width of the portion with \ion{Na}{1} age $t$ at $2v_\mathrm{out}t$. LASCO and HI1, however, cannot resolve the width of Phaethon's tail near perihelion: In one $t\sim$0.8~h photoionization lifetime at $r=0.14$~au, the tail broadens to only $2v_\mathrm{out}t\sim5000$~km, or ${\sim}7''$ from the closest distance $\varDelta=0.86$~au---below the $12''$ pixel scale of our highest resolution, LASCO C2 imagery. Rapid photoionization limits the broadening of \ion{Na}{1} tails of other sunskirting objects at lower $r$ to be even narrower.
\end{enumerate}

Now, we consider the impact of optical depth on tail profiles. A rigorous optically thick model with proper treatment of multiple scattering and particle dynamics would be far too computationally intensive for our analysis. Instead, we use a leading order approximation at low optical depths that neglects the impact of optical depth on the acceleration of \ion{Na}{1}. This simplification allows the use of the optically thin linear \ion{Na}{1} density $\Lambda$ to derive the optical depth and associated dimming along the tail.

The principal effect of optical depth is to reduce the number of scatterable solar \ion{Na}{1} D photons surviving into the tail before being scattered out of the tail by \ion{Na}{1}, and thus reduce the effective fluorescence efficiency of atoms in the tail. While scattered \ion{Na}{1} D photons can be rescattered by additional \ion{Na}{1} atoms on their way out of the tail, one scattered \ion{Na}{1} D photon will eventually escape the tail for every solar photon that excites an \ion{Na}{1} atom in the tail. The optical depth along the observer line of sight should therefore only minimally affect the observed tail brightness through anisotropic redirection of scattered \ion{Na}{1} D photons, favoring directions out of the tail with relatively lower optical depth, which we neglect in our model.

A solar photon can only be scattered by an \ion{Na}{1} atom if it has a wavelength $\lambda$ matching that of one of the \ion{Na}{1} D lines in the rest frame of the atom, neglecting the natural line shapes. In practice, the \ion{Na}{1} will have some nonzero velocity dispersion $\Delta v$, so any portion of the tail can scatter solar photons within a Doppler shift of $\pm\Delta v$. The $\dot{r}$ increases monotonically along the tail, so for each point along the tail with mean radial velocity $\dot{r}$, only the atoms directly sunward of that point with mean radial velocity between $\dot{r}-\Delta v$ and $\dot{r}$ contribute to the optical depth, corresponding to a segment of tail $L=\Delta v\times\dot{r}/\ddot{r}$ in length. Figure~\ref{fig:depth} illustrates this explanation.

We then model the tail as a series of these independent, cylindrical segments, where each segment corresponding to \ion{Na}{1} atoms of age $t$ has radius $v_\mathrm{out}t$, and therefore volumetric number density $N=\Lambda/(\pi v_\mathrm{out}^2 t^2)$. We define $\sigma_k\propto\Delta v^{-1}$ as the effective scattering cross section of \ion{Na}{1} atoms to scatterable photons within a $\pm\Delta v$ Doppler shift of the \ion{Na}{1} D$_k$ line, for which $\sigma_2\Delta v=(5.0\times10^{-13})$~m$^3$~s$^{-1}$ and $\sigma_1\Delta v=(2.5\times10^{-13})$~m$^3$~s$^{-1}$. The optical depth of the D$_k$ line for each segment is then

%\medskip
\begin{equation}
\begin{aligned}
\tau_k&=NL\sigma_k \\
&=\frac{\Lambda}{\pi v_\mathrm{out}^2 t^2}\times\frac{\Delta v\times\dot{r}}{\ddot{r}}\times\sigma_k \\
&=\frac{\Lambda\dot{r}}{\ddot{r}t^2}\times\frac{\sigma_k \Delta v}{\pi v_\mathrm{out}^2}
\end{aligned}
\end{equation}
%\bigskip

\noindent which we note does not actually depend on any particular choice of $\Delta v$, since $\Delta v$ is only included as part of the fixed factor $\sigma_k\Delta v$. We also note that $\tau_k\propto r^{-2}\propto T^{-1}$, so the optical depth is actually somewhat sensitive to our crudely estimated \ion{Na}{1} temperature, although the overall impact of this large uncertainty on the actual tail profile remains minimal at the low optical depths for which this model is valid.

Finally, we modulate the optical thin \ion{Na}{1} D$_k$ fluorescence efficiency along the tail by a factor of $(1-\exp(-\tau_k))/\tau_k$ to produce our final estimate for the tail brightness profile. We caution that since $\tau_k\to\infty$ as $t\to0$, even low $Q(\text{\ion{Na}{1}})$ sources can produce a section of optically thick tail that may not be well-reproduced by this model. However, our model suffices to confirm that Phaethon's observable \ion{Na}{1} tail is largely optically thin---with optical depth affecting its integrated brightness by up to only ${\sim}10\%$---which validates the physical parameters we fit under that assumption.

\bibliography{ms}

\end{CJK*}
\end{document}